\documentclass[fleqn,usenatbib]{mnras}

\usepackage{newtxtext,newtxmath}

\usepackage[T1]{fontenc}

\usepackage[hang,small,bf]{caption}
\usepackage[subrefformat=parens]{subcaption}
\usepackage{comment}
\usepackage{url}

\usepackage{color}
\usepackage[normalem]{ulem}
\usepackage{acronym}

\usepackage{graphicx}	

\newcommand{\msun}{\textnormal{M}_\odot}
\newcommand{\rsun}{\textnormal{R}_\odot}
\newcommand{\lsun}{\textnormal{L}_\odot}
\defcitealias{Hirai2018}{H18}

\acrodef{CCSN}{Core-collapse supernova}
\acrodefplural{CCSN}[CCSNe]{Core-collapse supernovae}
\acrodef{SN}{supernova}
\acrodefplural{SN}[SNe]{supernovae}
\acrodef{NS}{neutron star}
\acrodefplural{NS}[NSs]{neutron stars}
\acrodef{BH}{black hole}
\acrodefplural{BH}[BHs]{black holes}
\acrodef{ECI}{ejecta--companion interaction}
\acrodef{MS}{main-sequence}

\title[Observability of inflated companions after SNe]{Observability of inflated companion stars after supernovae in massive binaries}

\author[M.Ogata and R.Hirai]{
Misa Ogata,$^{1}$\thanks{E-mail: ogata@heap.phys.waseda.ac.jp},
Ryosuke Hirai$^{2,3,4}$,
Kotaro Hijikawa$^{5}$
\\
$^{1}$Advanced Research Institute for Science and Engineering, Waseda University,
3-4-1 Okubo, Shinjuku, Tokyo 169-8555, Japan\\
$^{2}$OzGrav: The Australian Research Council Centre of Excellence for Gravitational Wave Discovery, Clayton, VIC 3800, Australia\\
$^{3}$School of Physics and Astronomy, Monash University, VIC 3800, Australia\\
$^{4}$Department of Physics, University of Oxford, Keble Rd, Oxford, OX1 3RH, United Kingdom\\
$^{5}$Department of Astronomy, Graduate School of Science, The University of Tokyo, Bunkyo-ku, Tokyo, Japan
}
\date{Accepted XXX. Received YYY; in original form ZZZ}

\pubyear{2020}

\begin{document}
\label{firstpage}
\pagerange{\pageref{firstpage}--\pageref{lastpage}}
\maketitle

\begin{abstract}
We carry out a systematic study of the response of companion stars in massive binaries after being impacted by supernova ejecta. A total of 720 1D stellar evolution calculations are performed to follow the inflation and contraction of the star in response to the energy injection and how it depends on various parameters. We find that the maximum luminosity achieved during the inflated phase is only dependent on the stellar mass and we derive an analytic formula to describe the relation. There is also a tight correlation between the duration of expansion and the intersected energy. These correlations will be useful to constrain pre-supernova binary parameters from future detections of inflated companions. We also discuss the possible outcomes of the binary system when the companion inflation is taken into account. Based on simple binary population synthesis, we estimate that $\sim$1--3~\% of stripped-envelope supernovae may have observable inflated companions. Finally, we apply our models to the observed companion of SN~2006jc and place strong constraints on the possible pre-supernova binary parameters.


\end{abstract}

\begin{keywords}
binaries: close -- supernovae: general -- supernovae: individual: SN~2006jc -- stars: kinematics and dynamics
\end{keywords}



\section{Introduction}
    \acp{CCSN} are explosions that occur at the end of the lives of massive stars ($>8~\msun$), and they leave dense compact remnants such as \acp{NS} or \acp{BH}. The outermost regions of the progenitors usually consist of mostly hydrogen and helium which can be spectroscopically identified in \ac{SN} observations. However, a substantial fraction of \acp{CCSN} have much weaker or no trace of hydrogen, indicating a diversity in the structure of \ac{CCSN} progenitors \citep[]{Li2011, Smith2011}. Such hydrogen-deficient \acp{SN} are classified as type Ib/Ic/IIb depending on the spectroscopic features at the peak of the light curve, and are collectively called stripped-envelope \acp{SN} \citep[see][and references therein for details of the \ac{SN} classification]{Modjaz2019}. The progenitors of these stripped-envelope \acp{SN} lose their envelope during their evolution leading to \ac{SN}, through mass-loss mechanisms such as stellar winds or interactions with a binary companion.

    In binary systems, various binary interactions could change their evolution greatly. The specific form of interaction and their outcomes are sensitive to their initial separation and mass ratio. For example, the main mass-loss processes are Roche lobe overflow and common-envelope evolution. Through a combination of these interactions, stars can be stripped of their hydrogen envelopes \citep[e.g.][]{Podsiadlowski1992, Yoon2010, Yoon2015, Eldridge2008, Eldridge2017, Zapartas2019, Sravan2019} and some of them may even lose their helium layers prior to explosion \citep[e.g.][]{Tauris2013}. However because of the vast variety of possible evolutionary paths, the understanding of the whole picture is difficult.

    To understand binary evolution, the feedback from observations is essential. One example is the observation of the \ac{SN} themselves, which could be used to constrain the binary parameters at the point of SN explosion. Especially since one of the main channels for producing stripped-envelope \ac{SN} progenitors is through binary interactions, they would be good targets. Pre-SN observations of the \emph{progenitor} provide extra means to directly probe the properties of the progenitor \citep[]{Smartt2009, VanDyk2017}. For the case of stripped-envelope SNe, only several progenitors have been detected: type IIb SNe SN 1993J \citep[]{Aldering1994}, SN 2011dh \citep[]{Maund2011}, SN 2013df \citep[]{VanDyk2014}, SN 2008ax \citep[]{Folatelli2015}, SN 2016gkg \citep[]{Bersten2018}, and type Ib SNe iPTF13bvn \citep[]{Cao2013}, SN 2019yvr \citep[]{Kilpatrick2021}. Based on the pre-SN information, various binary evolution paths were advocated for individual cases \citep[e.g.][]{Bersten2014,Eldridge2015,Hirai2017,Sravan2018}. On the other hand, post-SN observations are also extremely important in confirming whether the progenitor had a binary companion or not. So far there are four stripped-envelope \acp{SN} with \emph{companion} detections, SN 1993J \citep{Maund2004}, SN 2001ig \citep[]{Ryder2018}, SN 2006jc \citep{Maund2016,Sun2019} and SN 2011dh \citep[]{Folatelli2014,Maund2019}, confirming that some stripped-envelope \ac{SN} progenitors are indeed produced through the binary channel. These companions are observed a few years after the explosion, when the luminosity of the \ac{SN} itself has faded sufficiently below that of the companion star.

    Moreover the observation of \ac{SN} in binaries help our understanding not only on the progenitors but also on the evolution after \ac{SN}. After the first direct detection of gravitational waves \citep[]{LIGO_BH}, compact binaries such as binary \acp{NS} or binary \acp{BH} have attracted wide attention. In particular, the detection of a binary \ac{NS} merger event was followed by observations of counterparts across the entire electromagnetic spectrum, providing valuable information for not only astrophysics, but also gravitational and nuclear physics \citep[]{LIGO_NS,LIGO_NS_astro}. Their evolutionary paths are not completely understood, but it is clear that the binary should have experienced and survived at least two \ac{SN} events. Each \ac{SN} has a finite probability of disrupting the system due to the natal kick imparted to the new-born \ac{NS}, strongly influencing the formation rate of binary \ac{NS} systems. \acp{SN} are also one of the most observable phases during the evolution of a massive binary system. Therefore, it is extremely important to understand the outcome of \acp{SN} in binary systems and establish ways to extract information of the binary properties from \ac{SN} observations.

    To study the nature of the binary companion after \acp{SN}, the effect of the interaction between \ac{SN} ejecta and the companion star (ejecta-companion interaction; ECI) has been studied for a long time through analytic modelling \citep[]{Wheeler1975, Meng2007}, numerical simulations \citep[e.g.][]{Fryxell1981, Marietta2000, Podsiadlowski2003, Pan2010, Pan2012, Liu2012, Shappee2013, Maeda2014, Zeng2020} and even laboratory experiments \citep{Domingo2019}. The application to \acp{CCSN} are still much more limited \citep{Hirai2014, Hirai2015, Liu2015, Rimoldi2016, Hirai2020}.

    \citet{Hirai2018} (hereafter \citetalias{Hirai2018}) carried out a large set of 2D hydrodynamical simulations of ECI with massive \ac{MS} companions ($10$--$20~\msun$). They found that the classical analytical model based on momentum conservation \citep[]{Wheeler1975} fails to reproduce their simulation results for the amount of unbound mass and imparted momentum. They constructed an alternative analytical model that can explain the imparted momentum along with the injected energy and how they are dependent on the stellar structure and orbital parameters. Moreover, they evolved the MS companion for 10,000 years after explosion, and found that the massive companion can become very luminous and inflated for up to tens of years due to the energy excess in the surface layers of the envelope.

    One example of the companion observation of \acp{SN} is SN 2006jc. SN 2006jc is a type Ibn SN which is one of the hydrogen-deficient SNe that have prominent narrow lines of He I in their spectra \citep[]{Foley2007}. An outburst was also observed in 2004 \citep[]{Pastorello2007}. The SN explosion was discovered by K. Itagaki on 2006 Oct 9 \citep{Nakano2006}, and occurred in the outskirt of UGC 4904. \citet{Maund2016} reported a late-time observation in 2010, and they discovered a faint blue source at the SN position. \citet{Sun2019} also report a late-time observation in 2017, and the magnitude in V-band was not so different between observations in 2010 and 2017. The stable optical brightness for over 7 years exclude other possibilities such as circumstellar matter interaction or light echoes, and conclude that the source is a stellar companion. Moreover the companion was inconsistent with being a \ac{MS} star and was instead located in the Hertzsprung gap that is known as an extremely short phase of stellar evolution. \citet{Sun2019} interpreted that the progenitor system is a binary whose mass ratio is close to unity and the companion has just recently evolved off its \ac{MS} phase.

    In this paper, we investigate the ECI in \acp{CCSN} in greater detail. \citetalias{Hirai2018} only followed the later evolution of these companions for a few models. They also did not refer to the post-SN evolution of the ``binary'' in detail. In this paper we aim to reveal the evolution of the systems after SN explosions using their formulation for much wider parameter regions and calculate the observability of such inflated companions. By studying the parameter dependence of the companion inflation, it would be useful for constraining pre-explosion binary parameters from follow-up observations of \acp{SN}.

    We review the results of \citetalias{Hirai2018} in Section~\ref{ECI:H18}, and discuss possible outcomes of the binary in the presence of expanding envelopes in Section~\ref{ECI:mode}. We will outline our calculation method and parameters in Section~\ref{sec:method}. We will show results of stellar modelling in Section~\ref{sec:result}. The classification of systems after SN and the observable distance will also be presented. In Section~\ref{sec:discussion}, we will present the rate that the expanding secondary is observed and the application to SN 2006jc. Implications of future detections of inflated companions will also be discussed. Finally we will summarize our work in Section~\ref{sec:summary}. Detailed calculation methods are shown in the Appendices \ref{app:orbit}, \ref{app:distance} and \ref{app:rate}.

\section{Evolution of Secondary Star with ECI}
\label{sec:ECI}

\subsection{Post-ECI inflation of the companion}
\label{ECI:H18}
    In this section we briefly review the model and results of \citetalias{Hirai2018}. They considered close binaries where the primary explodes as a CCSN with a massive MS companion. In systems with close separations, the primary should have experienced mass transfer prior to explosion and therefore should have lost most of the hydrogen envelope, i.e. the primary explosion is a stripped-envelope SN.

    Using the hydrodynamical code HORMONE \citep[]{Hirai2016}, they carried out 2D hydrodynamical simulations of SN blast waves colliding with their companions. As the ejecta hits the surface of the star, they form a forward shock that sweeps through the whole star and escapes from the other side. During this process, part of the kinetic energy of the ejecta is deposited into the envelope and drives it out of thermal equilibrium. There is hardly any mass stripped in this process, with only $\lesssim$1~\% of the mass lost even in the most extreme cases. They identified that most of the deposited energy is located near the surface layers and the excess energy declines as it goes deeper into the star. They also found that the total injected energy is $\sim$8--10~\% of the energy intersected by the secondary. This can be understood as a product of three factors; 1) a factor $(\gamma-1)/(\gamma+1)\sim1/4$, where $\gamma$ is the adiabatic index, which accounts for the energy thermalised at the bow shock that forms ahead of the star, 2) a geometrical factor 1/2 accounting for the deflection of ejecta at the stellar surface, 3) a factor $\sim$2/3 that accounts for the reduction in cross section due to stellar compression during the ejecta passage. Since this is a purely kinematic model, this should be applicable for any companion model as long as the stripped mass is negligible.

    They used the stellar evolution code MESA \citep[]{Paxton2010,Paxton2013,Paxton2015,Paxton2018,Paxton2019} to follow the secondary evolution for 10,000 years after ECI. They focused on a model in which the secondary mass is $M_2=10~\msun$, has a radius $R_2=5~\rsun$, and the explosion model with an ejecta mass $M_{\mathrm{ej}}=7.1~\msun$ and explosion energy $E_{\mathrm{expl}}=10^{51}~\mathrm{erg}$. The assumed metallicity was $Z=0.02$.

    In their hydrodynamical simulations, the excess energy is distributed uniformly around the envelope surface down to a certain radius and then decreases inverse proportionally to mass as it goes in. They created a fitting formula for the excess energy distribution $\Delta\varepsilon$,
    \begin{equation}
    	\Delta\varepsilon(m) = \frac{E_{\mathrm{heat}}}{m_h} \frac{{\mathrm{min}} \left(1, m_h / m \right)}{1+\ln \left(M_2 / m_h \right)},\label{eq:heat_dist}
    \end{equation}
    where $m$ is the mass coordinate from the surface. $E_{\mathrm{heat}}$ is the total injected energy which is estimated from the energy intersected by the secondary assuming the explosion is spherically symmetric,
    \begin{equation}
        E_{\mathrm{heat}} = p\times E_{\mathrm{expl}} \times \tilde{\Omega},\label{eq:totalheat}
    \end{equation}
    where $p\sim$8--10~\% is the energy injection efficiency. $\tilde{\Omega}$ is the fractional solid angle of the companion looking from the primary
    \begin{equation}
    	\tilde{\Omega} = \frac{\Omega}{4\pi} = \frac{1}{2} \left[ 1-\sqrt{1-\left( \frac{R_2}{a_\mathrm{i}} \right)^2} \right],
    \end{equation}
    where $a_\mathrm{i}$ is the pre-SN orbital separation. $m_h$ is the mass coordinate from the surface marking the transition from flat to decreasing energy excess, which can be estimated by
    \begin{equation}
    	m_h = \frac{\tilde{\Omega} M_{\mathrm{ej}}}{2}.\label{eq:mheat}
    \end{equation}
    Using these formulae, they inject energy into the companion star models and evolved them for 10,000 years.

    They obtained the time evolution of radius, and find that the secondary greatly expands and stays in the inflated state for several to tens of years. In some cases the maximum achieved radius far exceeds the initial orbital separation, meaning that the remnant of the primary could be embedded in the secondary envelope if the binary survives. This could give rise to various interesting phenomena, which will be discussed in the following section.

\subsection{Possible outcomes}
\label{ECI:mode}
Here we will discuss the possible outcomes of the system immediately after the first SN. The post-SN state can be classified depending on the post-SN orbital parameters and the degree of companion inflation due to ECI.

    We identify 6 states;
    \begin{itemize}
    	\item[-] state 1: the primary and secondary cores collide,
    	\item[-] state 2: the binary survives and the primary orbit is fully embedded in the inflated secondary envelope,
    	\item[-] state 3: the binary survives and the primary orbit is partially embedded in the inflated secondary envelope,
    	\item[-] state 4: the binary survives and the primary orbit does not intersect with the inflated secondary envelope at any point but could cause Roche lobe overflow,
    	\item[-] state 5: the binary survives in an orbit too wide to interact with each other,
    	\item[-] state 6: the binary is disrupted and evolve as two single stars.
    \end{itemize}

     State 1 occurs when the \ac{NS} natal kick is directed such that the post-SN periastron distance between the two stars is smaller than the original secondary star radius. In this case the primary NS is expected to rapidly inspiral within the secondary envelope and merge with each other. There could be some explosive phenomena as the NS accretes material in the envelope, leading to an unbinding of the outer layers and halting the spiral-in \citep[]{Fryer1996}. Contrarily, the NS may spiral in all the way to the centre, forming a Thorne-\.{Z}ytkow object \citep[T\.{Z}O;][]{Thorne1977, Leonard1994} which can have peculiar surface abundances such as enhanced lithium \citep[]{Podsiadlowski1995}. The ultimate fate of T\.ZOs are not known, but could lead to envelope collapse and \ac{BH} formation or explosions \citep[]{Podsiadlowski1995, Fryer1996, Moriya2018}.

\begin{figure}
    	\begin{minipage}{1.0\hsize}
    		\centering
    		\includegraphics[width=1.0\columnwidth]{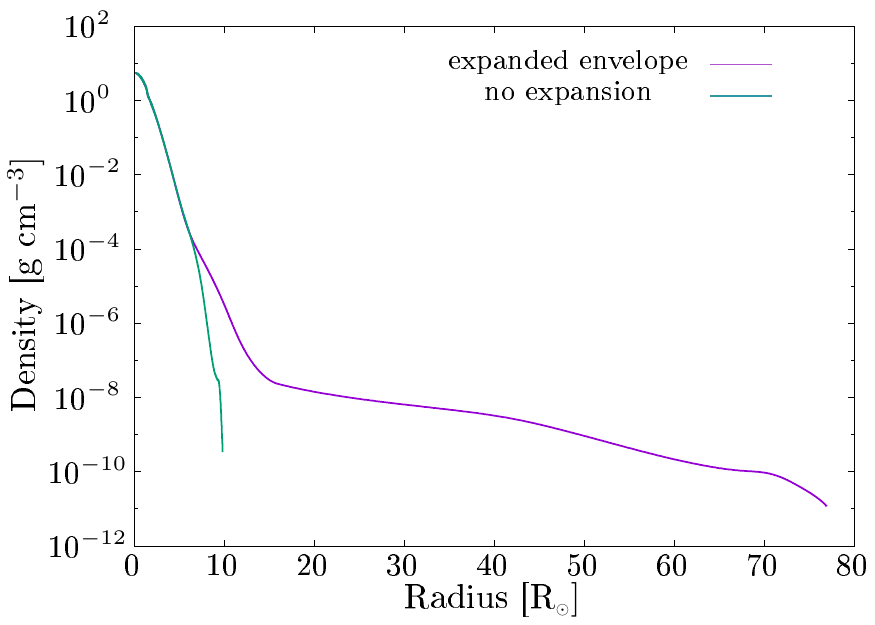}
    	\end{minipage}
    	\caption{Density distributions for the model with $M_2=15~\msun, R_2=7~\rsun, a_\mathrm{i}=40~\rsun, M_{\mathrm{ej}}=7.1~\msun, E_{\mathrm{expl}}=10^{51}\ \mathrm{erg}$. The purple curve is the density distribution of the inflated secondary and the green curve is the original profile.}
	    \label{density_dis}
    \end{figure}

     States 2 and 3 are when the binary survives with an orbit tight enough to be fully or partially engulfed by the inflated part of the secondary envelope. In Fig.~\ref{density_dis} we show an example of how the density structure inside the inflated part of the envelope can be. Only a very small amount of mass is heated and inflated, so the density in the region between the inflated and original surfaces is extremely low. This density is not sufficient to impart a meaningful drag force on the \ac{NS}, therefore not being able to affect the orbital evolution. However, there could still be efficient accretion onto the \ac{NS} that gives rise to some X-ray transients. It will only be observable if the outflow from the accretion is able to make its way out of the optically thick envelope. If the accretion feedback is significant, it may fully unbind the inflated layers of the envelope, enabling high energy radiation to escape and also suppresses the post-ECI inflation of the secondary. This means that even if the effect of ECI on the companion is the same, their appearances could be significantly different depending on whether the binary survived or not and what the post-SN orbit is.

State 4 is a case where the inflated secondary envelope can exceed its post-SN effective Roche lobe at periastron. It is possible that the secondary experiences Roche lobe overflow to the new-born \ac{NS}, creating a very brief X-ray binary phase. The stellar inflation may again be suppressed due to the mass transfer, although it is not clear how these inflated stars respond to mass-loss and the mass transfer process in eccentric binaries is still poorly understood.
Another possible outcome of state 4 is the formation of planets. \citet{Nakamura1991} proposed that if part of the inflated envelope can be transferred to the \ac{NS} and form a proto-planetary disk, it could explain the origin of pulsar planets.

State 5 is a trivial case where the binary survives but the post-ECI secondary does not inflate enough to exceed its Roche lobe. In such cases nothing particularly interesting would happen and would just continue its evolution as a regular binary.


In state 6, the binary disrupts and continues its evolution as two separate single stars. This case occurs when more than half of the total system mass is lost in the SN explosion, or the NS natal kick is large and directed in certain directions.

In this paper, we will explore on the observability of inflated companions after the first SN in the system. We will also take into account the post-SN binary properties as it may affect the maximum extent of the companion inflation. By studying the relative occurrence rates of each state, it can provide some insight into the observability of brief X-ray transients or pulsar planet formation.

\section{Numerical method}
\label{sec:method}

    Based on the methodology established in \citetalias{Hirai2018}, we extend the study of ECI and secondary expansion to a much wider parameter space. We  investigate the parameter dependence of the expansion radius of the secondary envelope and the inflated timescale. In this work we vary 5 model parameters, resulting in a total of 720 models. The relevant parameters are the companion mass $M_2$, the companion radius $R_2$, the binary separation $a_\mathrm{i}$, the ejecta mass $M_{\mathrm{ej}}$ and the explosion energy $E_{\mathrm{expl}}$. The combination of parameters used are summarized in Table \ref{initial}. The lower end of the secondary mass range is taken from 3~$\msun$, below which stars may not exist as companions to massive stars ($\gtrsim8~\msun$) because the star could not end its contraction phase within the massive stars' lifetime \citep[]{Neugent2020}. The largest secondary mass we use is $20~\msun$. Although it is possible that higher mass companion stars exist for \acp{CCSN}, such systems are less abundant so we ignore them in our study.
 We use the typical CCSN explosion energy of $10^{51}~\mathrm{erg}$, but we also consider hypernova-class energies of $10^{52}~\mathrm{erg}$ to see the energy dependence. The minimum orbital separations are chosen such that the secondary is close to Roche lobe filling, and the maximum separations are where the effect of ECI becomes negligible.

    For all our calculations we use the 1D stellar evolution code MESA (v10398). The initial stellar model is evolved up to the specified radius $R_2$, to account for the radius increase during the \ac{MS} phase. For the $M_2=3, 5~\msun$ models, we fix the radius to the zero-age main-sequence value as we do not expect much radius evolution within the lifetimes of CCSN progenitors. In real close binaries, the secondary should have slightly different structures from that of single stars due to mass transfer from the primary \citep[e.g.][]{Braun1995}, but we ignore this effect and use normal single MS models.

    To each stellar model, we artificially inject energy following Eq.~(\ref{eq:heat_dist})--(\ref{eq:mheat}), over a timescale of $\sim$1~yr. For the lower mass models ($M_2=3, 5~\msun$), we choose a slightly longer injection time of $\sim$3--10~yr to avoid numerical difficulties due to the longer thermal timescale of the star. All injection timescales are set to be significantly shorter than the thermal timescale, and we confirmed that the timescale does not have a huge effect on the evolution as long as this condition is satisfied. The true energy injection timescale is much shorter ($\sim$few~hr, corresponding to the shock crossing time), and therefore we define the time since the end of the artificial heating phase as the time since SN. We use a fixed energy injection efficiency $p=0.08$\footnote{Strictly speaking, this is dependent on the stellar structure and orbital separation, where $p$ increases as the orbital separation increases \citepalias{Hirai2018}. However, the variation is small ($p\sim$0.08--0.1) so we fix the value for simplicity.}. A fixed metallicity $Z=0.02$ is used, however we run a few additional models to explore the metallicity dependence. We use the method adopted in \citet[]{Henyey1965} with the mixing length parameter set to $\alpha_\textsc{mlt}=2.0$ for convection, and the so-called MLT++ feature is switched off \citep[Section 7.2 in][]{Paxton2013}. We apply the ``Dutch wind'' scheme \citep[see][]{Glebbeek2009} throughout, although the effect of winds is negligible for our applications. Simple grey atmosphere boundary conditions are applied to the outer boundary. The MESA \texttt{inlist} and \texttt{run\_star\_extras.f} files for the calculations are made publicly available in Zenodo \footnote{\url{https://zenodo.org/record/4624586##.YHH6Cq_7Q2w}}. The models are run for 10,000~yr, which encases the whole expansion and contraction phase after energy injection.


    \begin{table}
      \centering
      \begin{tabular}{ccccc}     \hline
        $M_2 [\msun]$ & $R_2 [\rsun]$ & $a_\mathrm{i} [\rsun]$ & $M_{\mathrm{ej}} [\msun]$ & $E_{\mathrm{expl}} [\mathrm{erg}]$  \\  \hline
        3 & 2 & 10, 20, 30, 40, 60 & 2, 3, 5 & $10^{51},10^{52}$  \\
        5 & 2.7 & 10, 20, 30, 40, 60 & 2, 3, 5 & $10^{51},10^{52}$ \\
        10 & 6, 7, 8, 9, 10 & 20, 30, 40, 60 & 2, 3, 5, 7.1, 10 & $10^{51},10^{52}$  \\
        15 & 6, 7, 8, 9, 10 & 20, 30, 40, 60 & 2, 3, 5, 7.1, 10 & $10^{51},10^{52}$  \\
        20 & 6, 7, 8, 9, 10 & 20, 30, 40, 60 & 2, 3, 5, 7.1, 10 & $10^{51},10^{52}$  \\ \hline
      \end{tabular}
      \caption{The parameter space explored in our study. The stellar radius for $M_2=3, 5~\msun$ are fixed to zero-age main-sequence values. The total number of models are 720.}
      \label{initial}
    \end{table}

\begin{figure}
    	\begin{minipage}{1.0\hsize}
    		\centering
    		\includegraphics[width=0.85\columnwidth]{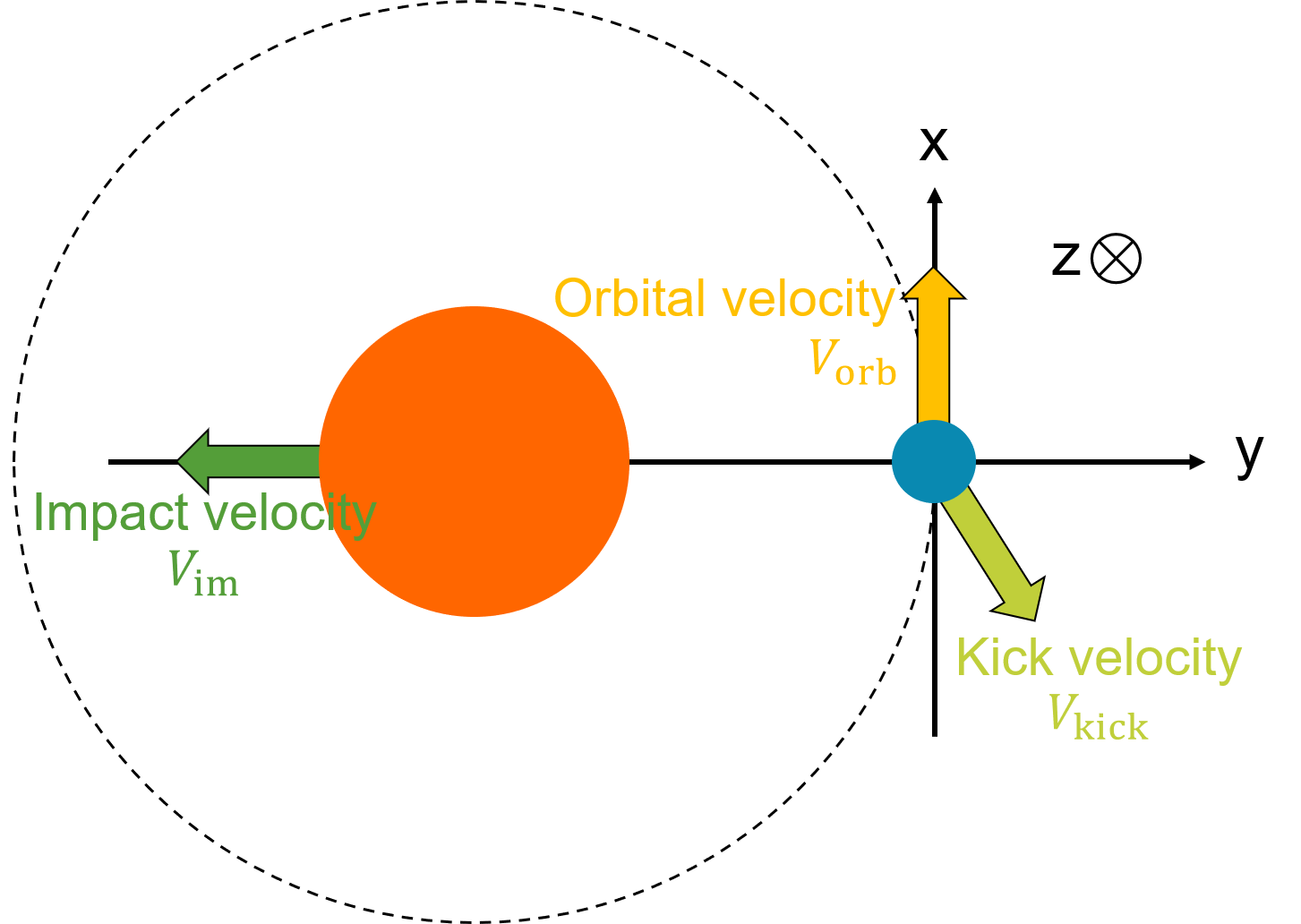}
    	\end{minipage}\\
    	\begin{minipage}{1.0\hsize}
    		\centering
    		\includegraphics[width=0.65\columnwidth]{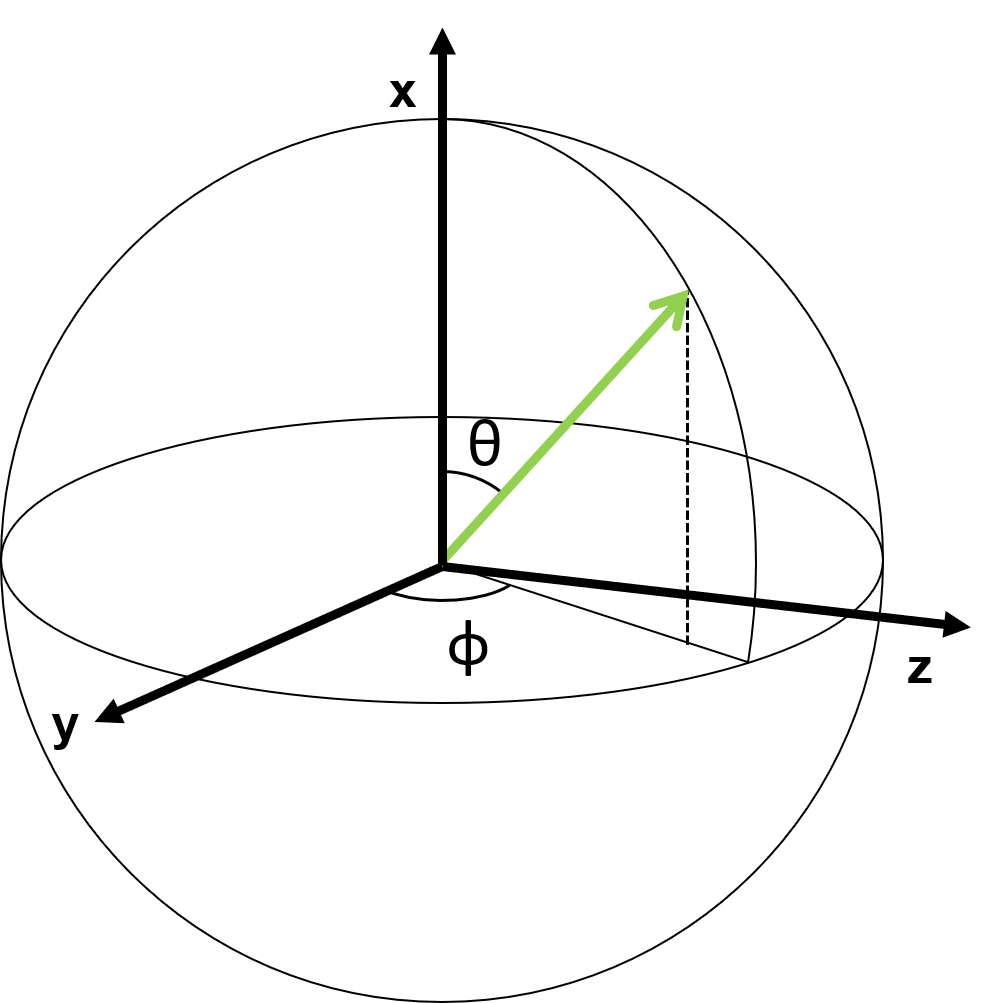}
    	\end{minipage}
    	\caption{Schematic diagrams of the coordinate system used for the orbit calculations. (Upper panel) the red large circle expresses the secondary MS star and the blue small circle is the primary. The three arrows represent the orbital velocity (orange), the impact velocity (green) and the kick velocity (light green). The dotted circle expresses the circular orbit before explosion. (Lower panel) the definition of angles for the kick velocity. The x- and y-axis is the same as the upper panel.}
    	\label{setup}
    \end{figure}

    We also consider the binary properties immediately after the SN. It is determined by the amount of mass loss from the system, the NS natal kick and the impact velocity which is defined as the velocity given to the companion due to the momentum imparted by the ejecta. We assume initially circular orbits for all models. This is a reasonable assumption in the relevant parameter regions because the orbit should have been tidally circularised at close separations. We analytically calculate the post-SN orbital parameters by assigning a kick velocity and direction. In Fig.~\ref{setup} we show the coordinate system we use for the orbit calculations. We set the origin at the centre of the primary star (exploding star), with the x-axis in the direction of the initial orbital velocity $V_i$ and the y-axis directed from the companion to the primary. The z-axis is antiparallel to the orbital angular momentum. We also define angles $\theta$ and $\phi$ as shown in the lower panel of Fig.~\ref{setup}. The impact velocity always points in the negative y-direction whereas the natal kick direction is arbitrary. Details of how we calculate the post-SN orbital separation $a_\mathrm{f}$, eccentricity $e$ and Roche-lobe radius $R_\mathrm{rl}$ are described in Appendix~\ref{app:orbit}.


    The outcome of the SN can be classified by comparing the orbital parameters and the inflated envelope radius,
    \begin{itemize}
    	\item[-] state 1: $e<1\wedge a_\mathrm{f}(1-e)<R_2$,
    	\item[-] state 2: $e<1\wedge a_\mathrm{f}(1+e)<R_{\mathrm{2,max}}$,
    	\item[-] state 3: $e<1\wedge a_\mathrm{f}(1-e)<R_{\mathrm{2,max}}$,
    	\item[-] state 4: $e<1\wedge R_{\mathrm{rl,f}}<R_{\mathrm{2,max}}$,
    	\item[-] state 5: $e<1\wedge R_{\mathrm{rl,f}}>R_{\mathrm{2,max}}$,
    	\item[-] state 6: $e>1$,
    \end{itemize}
    where $R_{\mathrm{rl,f}}, R_{\mathrm{2,max}}$ are the post-SN effective Roche-lobe radius at periastron and the maximum radius of the secondary during the inflated phase.


    Our methodology is based on various assumptions. First, we have assumed that the explosion is spherically symmetric. In reality, CCSNe are known to be rather aspherical, and therefore the energy intersected by the secondary can vary depending on the degree of asphericity. However, this uncertainty will be masked by the wide range in possible explosion energies, so we do not take the asphericity into account. Another assumption is that the ejecta velocity is much faster than the orbital velocity. This only breaks down in extremely tight orbits such as for ultra-stripped SNe \citep[see e.g.][]{Tauris2013}. In such situations, the mass loss can no longer be treated as an impulsive event and hence the orbital response will behave differently \citep[]{Fryer2015}. We do not deal with ultra-stripped SNe in this paper, so the impulsive approximation is always valid. We also assume that there is no mass stripping by the ECI. This is a good approximation for MS companions where the unbound masses are $\lesssim$~1~\% \citep[]{Liu2015, Rimoldi2016, Hirai2018}. This is invalid for more evolved companions such as Hertzsprung gap stars or red supergiants where there can be significant mass loss \citep[]{Hirai2014, Hirai2020}, but these cases are rare and we ignore such systems in this paper.

\section{Results}
\label{sec:result}

\subsection{Time evolution of stellar properties}
\label{res:radHR}

\begin{figure}
    	\centering
    	\includegraphics[width=1.0\columnwidth]{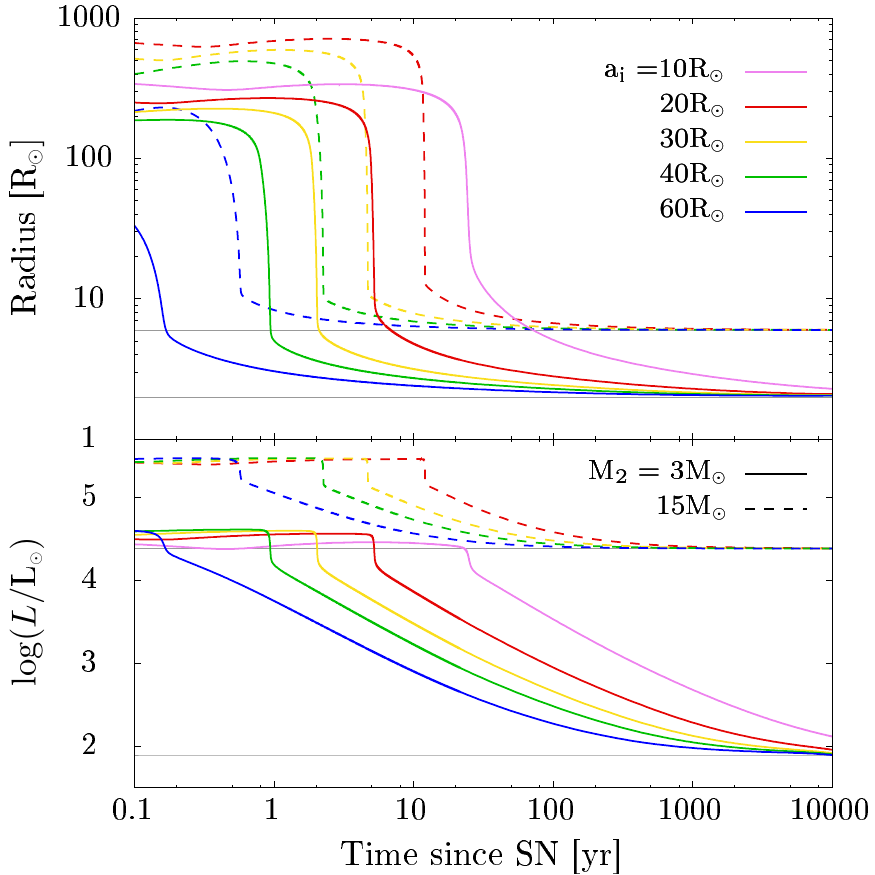}
    	\caption{Time evolution of the secondary radius (upper panel) and luminosity (lower panel) after energy injection for the models with $M_2=3~\msun, R_2=2~\rsun, M_{\mathrm{ej}}=2~\msun, E_{\mathrm{expl}}=10^{51}\ \mathrm{erg}$ (solid curves), and $M_2=15~\msun, R_2=6~\rsun, M_{\mathrm{ej}}=2~\msun, E_{\mathrm{expl}}=10^{51}\ \mathrm{erg}$ (dashed curves). Different colours denote different orbital separations $a_i=10$ (only for $M_2=$3~$\msun$), 20, 30, 40, 60~$\rsun$. The grey solid lines mark the original stellar radii $R_2$ and the original stellar luminosities. The time is measured from the end of the artificial heating.}
        \label{tr}
    \end{figure}

    In the upper panel of Fig.~\ref{tr} we show the time evolution of radius for the $M_2=3~\msun$ and $15~\msun$ ($M_{\mathrm{ej}}=2~\msun, E_{\mathrm{expl}}=10^{51}~\mathrm{erg}$) models.
The initial radius for the $M_2=15~\msun$ models are $ R_2=6~\rsun$, and the separations are varied between $20~\rsun$ and $60~\rsun$. Most curves have qualitatively the same shape, where the radius stays roughly constant and then abruptly shrinks back to its original radius after $\sim$1--20~yr. 
The star is still quite overluminous for $\sim$100--10,000~yr even after the radius has retracted. However, the temperature is much hotter once the star has contracted, so the drop in luminosity at longer wavelengths should be more prominent than the shape in this figure.

For a fixed stellar mass, it is clear that the expansion radius and timescale increase with decreasing orbital separation. Especially, the $a_\mathrm{i}<20~\rsun$ models expand largely for more than 10 years, which may be long enough to be observable as the SN itself fades away.
For fixed orbital separations, the expansion timescales are shorter for larger secondary mass. This is because the luminosity of the inflated stars are larger and therefore takes less time to radiate away the excess energy. Although we only show the time evolution for select systems, the overall shape and parameter dependences are similar for all other models.

\begin{figure*}
    	\centering
    	\includegraphics[width=1.9\columnwidth]{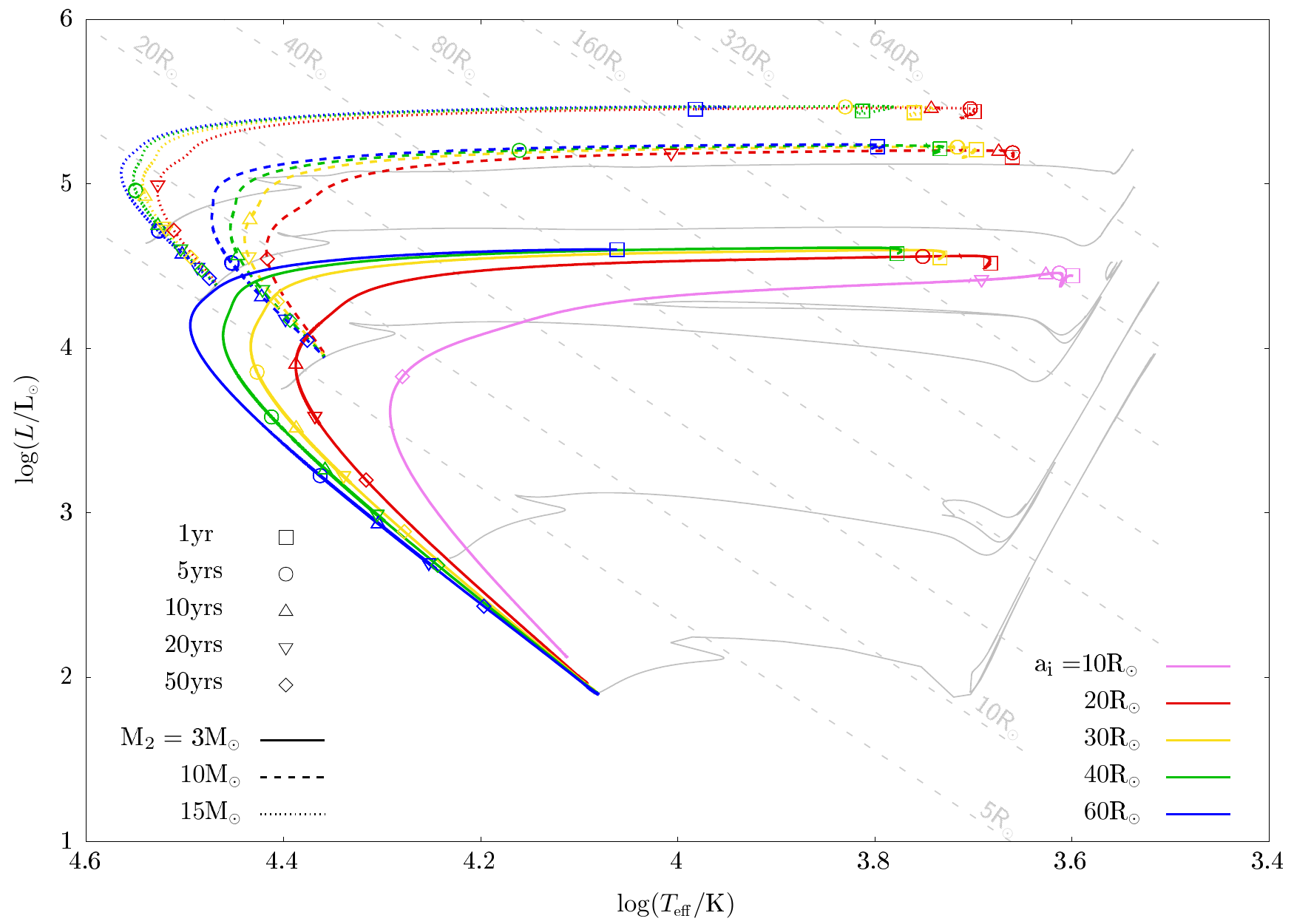}
    	\caption{HR-diagram after explosion for the models $(M_2,R_2)=(3,2), (10,6), (15,6), M_{\mathrm{ej}}=2~\msun, E_{\mathrm{expl}}=10^{51}\ \mathrm{erg}$. Grey solid curves are the normal evolutionary tracks of $3, 5, 10, 15, 20~\msun$ stars. Grey dashed lines are equal radius lines of $5, 10, 20, 40, 80, 160, 320, 640~\rsun$. The squares, circle and triangles are time stamps at $1, 5, 10, 20, $ and $50$ years after SN explosion.}
        \label{tHR}
    \end{figure*}

    In Fig. \ref{tHR} we show the evolution on the HR-diagram for the models $(M_2,R_2) = (3,2), (10,6), (15,6)$, $M_{\mathrm{ej}}=2~\msun, E_{\mathrm{expl}}=10^{51}~\mathrm{erg}$.
    As the envelope expands, the secondary jumps to the cooler and more luminous region in the HR-diagram and stays there for several years. Some models inflate close to the Hayashi limit, but do not fully reach it. After that it rapidly returns to the original state as the envelope contracts. The maximum radius is larger for shorter separations, therefore the closer models reach lower temperatures. On the other hand, the maximum luminosity is almost independent of any parameters other than initial mass.

\begin{figure}
    	\centering
    	\includegraphics[width=1.0\columnwidth]{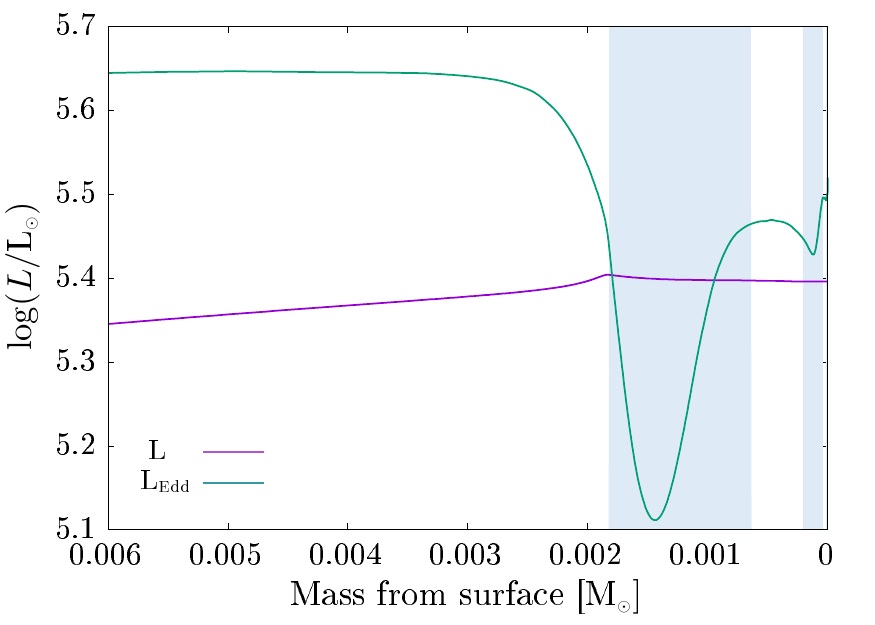}
    	\caption{Luminosity distribution in mass coordinate near the surface for the model with $M_2=15~\msun, R_2=9~\rsun, a_\mathrm{i}=60~\rsun, M_{\mathrm{ej}}=7.1~\msun, E_{\mathrm{expl}}=10^{51}\ \mathrm{erg}$ at the point of maximum expansion. The purple and green curves are the stellar luminosity and the Eddington luminosity respectively. The blue shaded region corresponds to the convective region.}
        \label{L_surf}
    \end{figure}

\begin{figure}
    	\centering
    	\includegraphics[width=1.0\columnwidth]{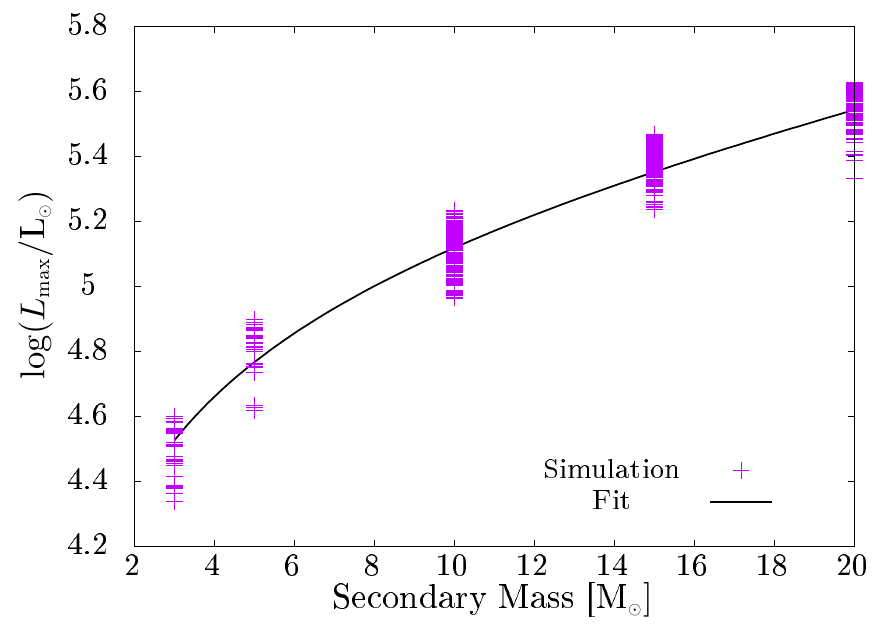}
    	\caption{Maximum luminosity as a function of secondary mass for our 720 models. The black curve is the Eddington luminosity computed with Eq.~(\ref{Lmax}).}
        \label{Lmax}
    \end{figure}

    To demonstrate why the maximum luminosity is roughly constant for a given stellar mass, we show the luminosity distribution near the surface of the star during the inflated phase in Fig. \ref{L_surf}. The luminosity distribution increases monotonically in the star because of the way the excess energy is injected (Eq.~(\ref{eq:heat_dist})). It then flattens out beyond some radius very close to the surface ($M_2-m\lesssim0.002~\msun$) above which most of the envelope is convective and the energy is efficiently transported. The bottom of this convective region corresponds to where the luminosity intersects with the local Eddington luminosity. At this specific layer, the local Eddington luminosity is determined by the Fe opacity bump (slightly off the peak) which is located at temperature regions of $\log(T/\mathrm{K})\sim$5.2--5.4. We find a similar structure for all of our models, indicating that the maximum luminosity after ECI is always determined by the Eddington luminosity evaluated by the Fe opacity. Because the Fe opacity is sensitive to the temperature and density, the actual opacity value at this layer is different for different mass models but within a reasonably small range. For each stellar mass, the average value of the opacity at the bottom of the convective layer is $\kappa\sim$1.25, 1.00, 1.03, 0.85, 0.76~cm$^2$~g$^{-1}$ for the $M_2=3, 5, 10, 15, 20~\msun$ models respectively. The average opacities decrease linearly as the secondary mass increases, and can be fit with the function
    \begin{equation}
     \kappa_\mathrm{fit} = 1.24~\mathrm{cm}^2 \mathrm{g}^{-1}\left(1-0.02 \frac{M_2}{\msun}\right).
    	\label{opacity}
    \end{equation}
This is consistent with the fact that at a layer with a given temperature, the density decreases as the mass of the star increases so the opacity decreases too. Using this fitting formula, we can analytically estimate the maximum post-ECI luminosity as
\begin{equation}
 L_{\mathrm{max,ana}}=\frac{4\pi GM_2c}{\kappa_\mathrm{fit}},
 \label{eq:Lmax}
\end{equation}
where $G$ is the gravitational constant and $c$ is the speed of light.
 We compare this against the simulated results for $L_{\mathrm{max}}$ in Fig.~\ref{Lmax}. The analytical estimates are in good agreement with the simulated results.

\subsection{Systematic study}
\label{res:allmodel}

    In our simulations, the response of the companion after ECI is almost like a phase transition between the inflated and original states (Fig.~\ref{tr}), and therefore can be characterised by just three variables: the maximum luminosity, maximum radius and duration of expansion. We have already shown that the maximum luminosity is only dependent on the stellar mass. In this section we will study the maximum radius and the duration of expansion and how they depend on the other input parameters. 

\begin{figure}
    	\centering
    	\includegraphics[width=1.0\columnwidth]{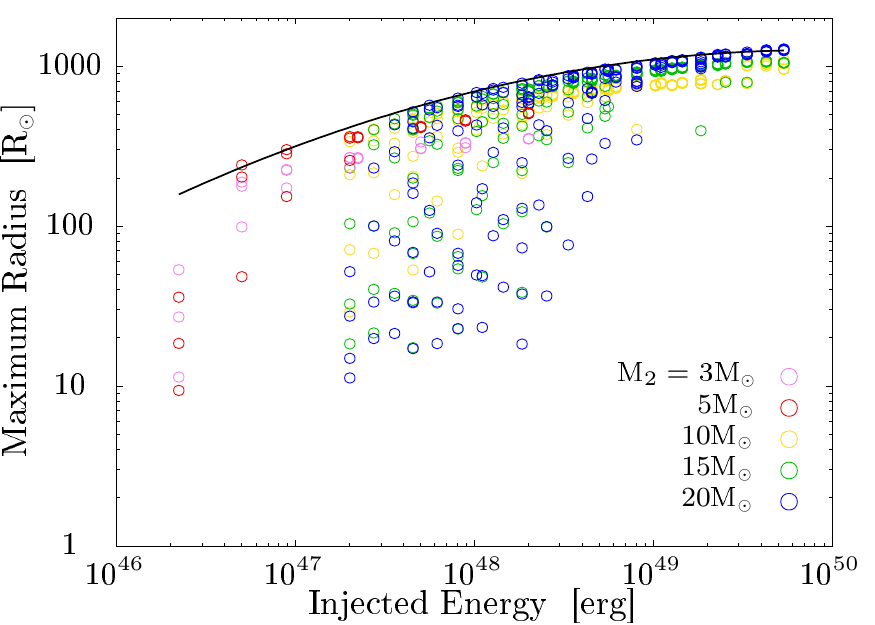}
    	\caption{Relation between the energy intersected by the companion and the maximum inflated radius. Open circles are simulated results and the black curve is an empirical fit based on Eq.~(\ref{eq:rmax_fit}). Colours of the circles denote the secondary mass. }
        \label{Rmax}
    \end{figure}

    The relation between the injected energy as defined in Eq.~(\ref{eq:totalheat}) and the maximum radius is presented in Fig.~\ref{Rmax}. There is a relatively strong correlation between the injected energy and maximum radius especially at the higher energy end. Models that fall off the relation are the models with smaller initial radius, and the expansion radius become small with the same injected energy due to the higher specific binding energy of the envelope. We empirically fit a curve to the maximum radii with the function
    \begin{equation}
        \log_{10} \frac{R_{\mathrm{max}}}{\rsun} = -\frac{1}{14}\left( \log_{10} \frac{E_{\mathrm{heat}}}{0.08\times10^{51}~\mathrm{erg}} \right)^2 + 3.1,
	 \label{eq:rmax_fit}
    \end{equation}
which is shown as the black curve.

\begin{figure}
    	\centering
    	\includegraphics[width=1.0\columnwidth]{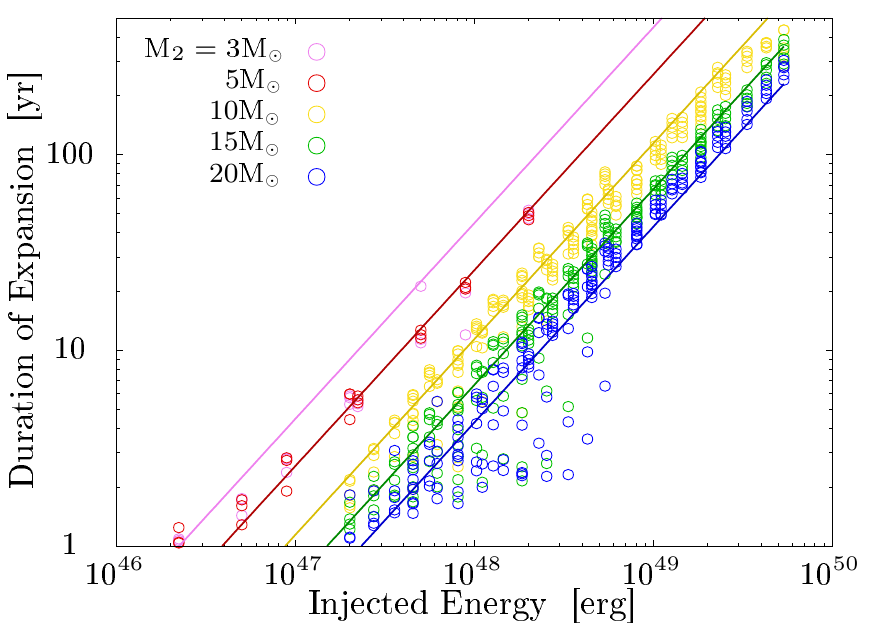}
    	\caption{Relation between the energy intersected by the companion and the duration of expansion. Open circles are simulated results and solid lines are analytical fits based on Eq.~(\ref{eq:t_inf_fit}). Colours of the circles and lines denote the secondary mass.}
        \label{timescale}
    \end{figure}

    In Fig.~\ref{timescale}, we display the duration of expansion as a function of the injected energy. The duration of expansion is defined as the time from the energy injection up to when the radius shrinks to half of the maximum radius. There is a remarkably tight correlation that can be expressed as power-laws and have different coefficients according to the secondary mass. More massive secondaries have shorter inflated timescales even if the injected energy is the same. Solid lines represent analytical fits to the correlation with the function
    \begin{equation}
     \tau_\mathrm{inf}=\alpha\frac{E_\mathrm{heat}}{L_\mathrm{max,ana}}
     \label{eq:t_inf_fit}
    \end{equation}
    where $L_\mathrm{max,ana}$ is computed from Eq.~(\ref{eq:Lmax}) and $\alpha$ is a fitting coefficient which we set to $\alpha=0.181$. Although in principle $\alpha$ can be mass-dependent, we find that the analytic curves fit the simulated results fairly well with a universal value. This implies that the inflated state of the star is only maintained until $\sim$18.1~\% of the injected energy is radiated away. We have compared this timescale with the thermal timescale of the heated part of the envelope ($m_h$ from the surface), but found no clear correlation, implying that $m_h$ does not directly determine the depth of the inflated part of the envelope. The dependence on other parameters such as ejecta mass or initial stellar radius seem to be much weaker.

\begin{figure}
    	\centering
    	\includegraphics[width=1.0\columnwidth]{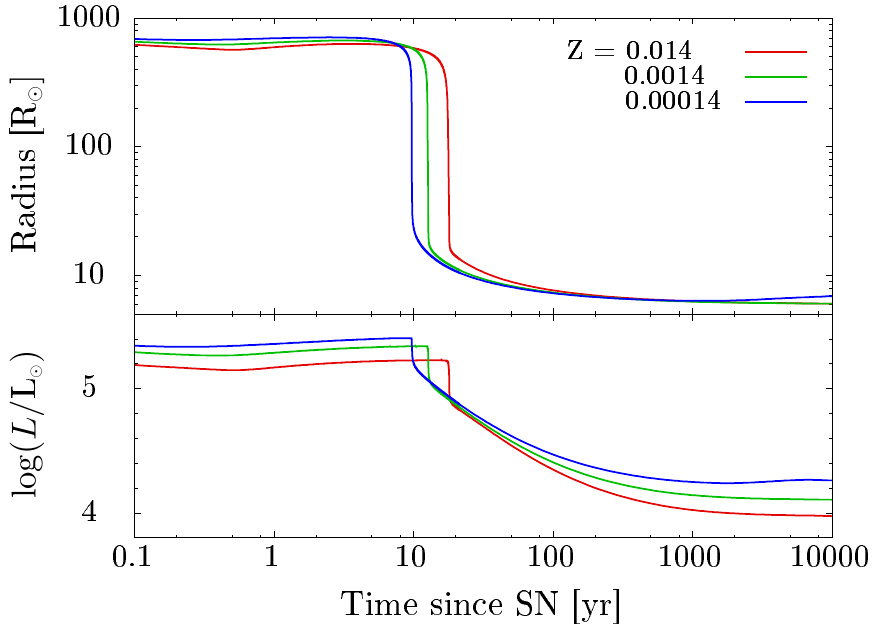}
    	\caption{Time evolution of the secondary radius (upper) and luminosity (lower) after energy injection with different metallicities. Model parameters are $M_2=10~\msun, R_2=6~\rsun, a_i=20~\rsun, M_{\mathrm{ej}}=2~\msun, E_{\mathrm{expl}}=10^{51}~\mathrm{erg}$. Different colours correspond to metallicities of $Z=0.014, 0.0014, 0.00014$.}
        \label{metallicity}
    \end{figure}

It should be noted that the maximum radius and luminosity can be rather sensitive to the choices made for our calculations. Because the maximum luminosity is restricted by the Eddington luminosity determined around the iron bump, it is quite sensitive to the metallicity of the star. We compared the evolution with different metallicities in Fig.~\ref{metallicity} and found that the maximum luminosity can increase by $\sim60\%$ when the metallicity is reduced by two orders of magnitude. The inflated radius increases and the duration of expansion decreases correspondingly. We will postpone a more thorough investigation of metallicity effects to future work.

The maximum radius can particularly be dependent on the treatment of the outer boundary condition. As the large amount of energy excess makes its way towards the surface, it could develop dynamical instabilities \citep[e.g.][]{Shaviv1999}, which cannot be correctly accounted for in our MESA simulations. Such effects can possibly drive super-Eddington winds \citep[]{Owocki2004} or dynamically eject the outer layers. It is also known that dynamical treatments of the outer boundary can change the stellar radius considerably compared to the usual hydrostatic treatments \citep[e.g.][]{Poniatowski2021}. Properly accounting for these effects will require a full hydrodynamical treatment in multi-dimensions, which is out of the scope of this paper. However, we expect the maximum luminosity and inflated timescales are much less sensitive to the numerical choices and the qualitative behaviour of the maximum radii should still hold, i.e. larger energy injection should lead to larger inflation.

\subsection{The fate of the system after ECI}
\label{res:mode}

\begin{figure}
    	\centering
        \includegraphics[width=1.0\columnwidth]{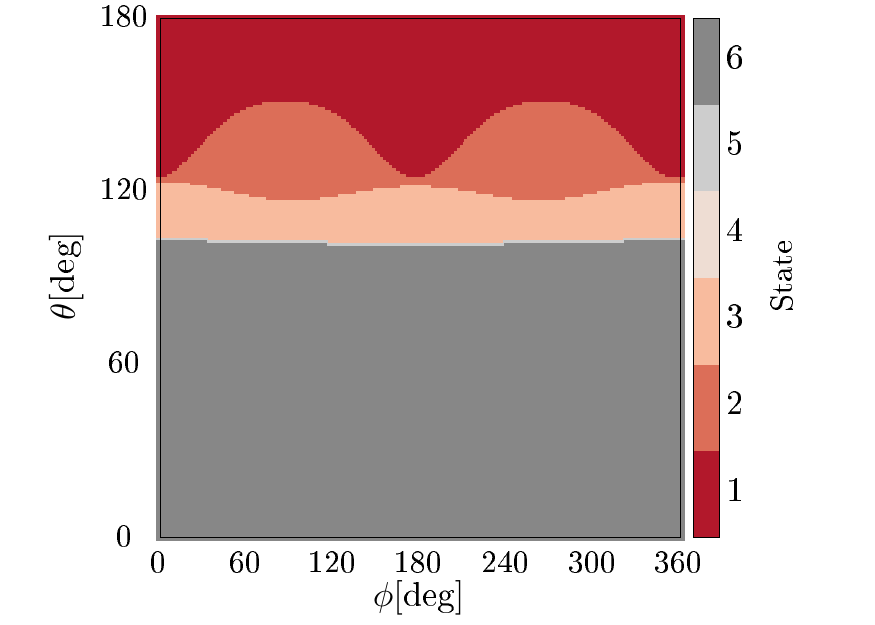}
	    \caption{Example system state map after explosion, colour-coded according to the kick direction. The binary model parameters are $M_2=15~\msun, R_2=7~\rsun, a_\mathrm{i}=40~\rsun$ and the kick velocity is fixed to $v_{\mathrm{kick}}=300~\mathrm{km~s^{-1}}$. The two angles $\theta, \phi$ for the kick direction are defined in Fig.~\ref{setup}, and the states are defined in Section~\ref{ECI:mode}: (1) primary and secondary collide, (2) binary survives and the primary orbit is fully embedded in the secondary envelope, (3) binary survives and the primary orbit is partially embedded in the secondary envelope, (4) binary survives and secondary envelope exceeds its Roche lobe at periastron, (5) binary survives in a wide orbit where there is no immediate interaction and (6) binary is disrupted.}
	    \label{map}
    \end{figure}

    To investigate the post-SN state of the system, we computed the post-explosion orbital parameters. The orbital separation and the eccentricity depend on the magnitude and direction of the natal kick imparted to the new-born NS. We show in Fig.~\ref{map} the states of the systems after explosion for a given kick velocity for the model with $M_2=15~\msun, R_2=6~\rsun, a_\mathrm{i}=60~\rsun, M_{\mathrm{ej}}=7.1~\msun, E_{\mathrm{expl}}=10^{51}~\mathrm{erg}, V_{\mathrm{kick}}=300~\mathrm{km~s^{-1}}$. The inflated secondary reached a radius of $R_\mathrm{max}=86~\rsun$ in this model. It illustrates how the six outcomes classified in section \ref{ECI:mode} appear depending on the kick direction.
    In the case that the kick velocity is pointed opposite to the orbital motion ($\theta\sim180^{\circ}$), the relative velocity between the primary and the secondary decreases and the two stars are destined to collide (State 1). At slightly more moderate angles ($150^\circ\lesssim\theta\lesssim120^\circ$), the stars can avoid colliding and instead the full orbit is embedded in the inflated part of the secondary envelope depending on $\phi$ (state 2). In the region $100^\circ\lesssim\theta\lesssim120^\circ$, the orbit widens after the SN and becomes partially embedded in the inflated secondary envelope (state 3). These cases are potentially interesting as they may cause some brief X-ray transients as they plunge in and out of the inflated envelope.
As the kick angle decreases and becomes closer to being aligned with the orbit ($\theta\sim0^{\circ}$), the relative velocity increases and they will simply become unbound and evolve as two single stars.

\begin{figure}
    	\begin{minipage}{1.0\hsize}
    		\centering
    		\includegraphics[width=0.9\columnwidth]{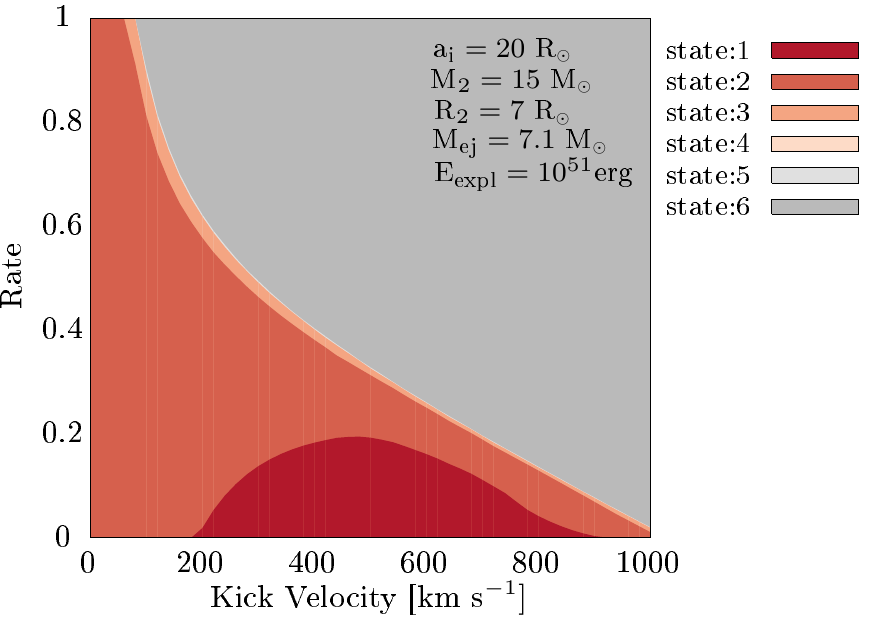}
	    \end{minipage}\\
	    \begin{minipage}{1.0\hsize}
	    	\centering
	    	\includegraphics[width=0.9\columnwidth]{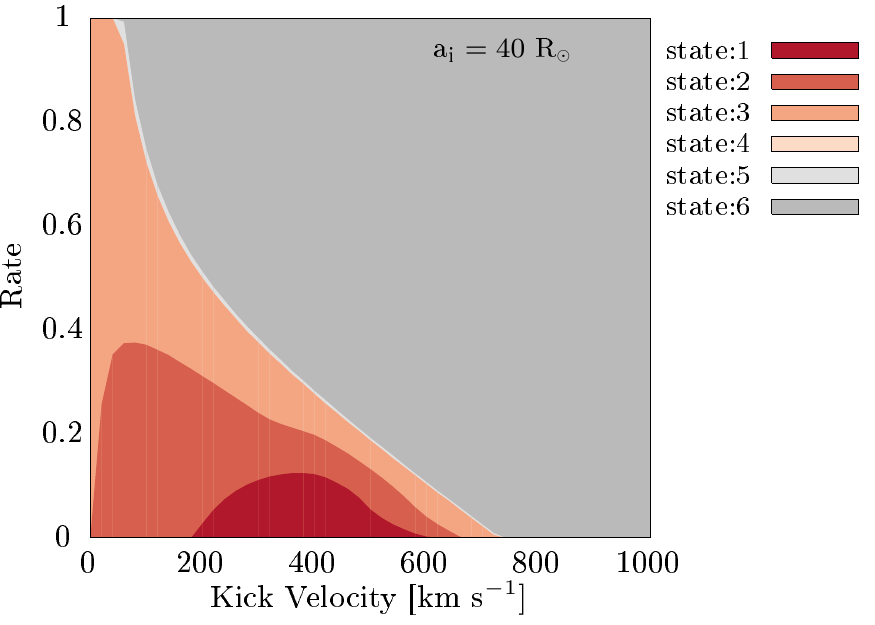}
	    \end{minipage}\\
	    \begin{minipage}{1.0\hsize}
	    	\centering
	    	\includegraphics[width=0.9\columnwidth]{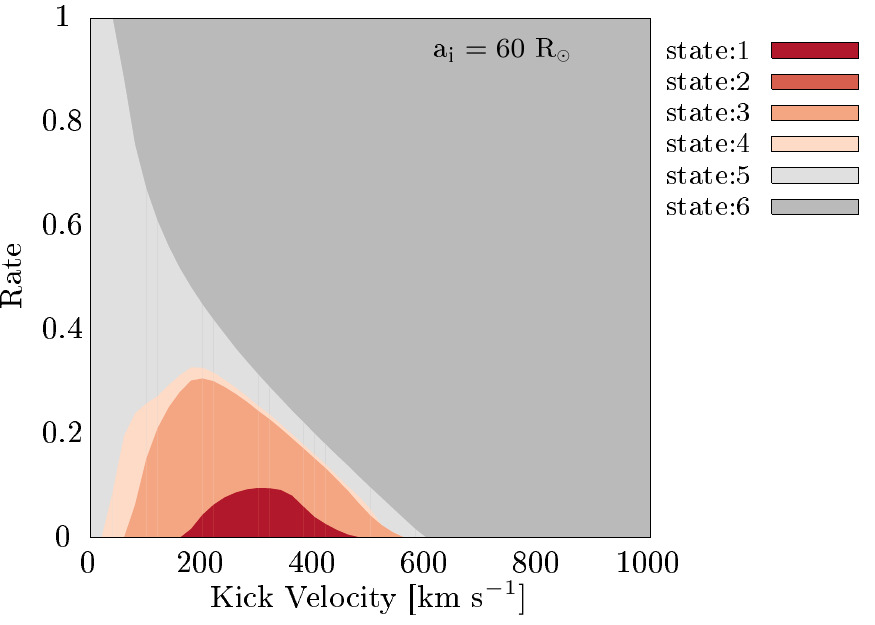}
	    \end{minipage}
	    \caption{System state rates as a function of kick velocity for the binary model $M_2=15~\msun, R_2=7~\rsun, M_{\mathrm{ej}}=7.1~\msun, E_{\mathrm{expl}}=10^{51}\ \mathrm{erg}$. Separations are $a_\mathrm{i}=20, 40, 60~\rsun$ from top to bottom panels respectively. The states are defined in Section~\ref{ECI:mode}. The corresponding colours are the same as in Fig.~\ref{map}.}
	    \label{rate_dis}
    \end{figure}

    The pattern in Fig.~\ref{map} can drastically change depending on the magnitude of the kick velocity. In Fig.\ref{rate_dis}, we show the probability of each outcome as a function of the kick velocity. For each given kick velocity, the probilities are computed by integrating the area in Fig.~\ref{map} assuming the kick direction is isotropically distributed. We focus on the same binary model with $M_2=15~\msun, R_2=7~\rsun, M_{\mathrm{ej}}=7.1~\msun, E_{\mathrm{expl}}=10^{51}~\mathrm{erg}$ and the three panels display results for separations $a_\mathrm{i}=20, 40, 60~\rsun$. The maximum inflated radii are $R_\mathrm{max}=346,86,40~\rsun$ for the $a_\mathrm{i}=20, 40, 60~\rsun$ models respectively. In all panels the fraction of binary disruption (state 6) rapidly increases with the kick velocity. In cases where the system keeps its binarity, the exact configuration is strongly dependent on both the kick velocity and initial orbital separation. The primary is more frequently embedded in the companion envelope (state 2) in the models with closer separation. This is because the secondary expansion is greater for smaller separation as seen in Fig.~\ref{tr} and satisfies $R_\mathrm{max}\geq a_\mathrm{i}$. Due to the fact that the post-SN periastron distance cannot exceed the original separation regardless of the kick velocity, part of the orbit will always intersect with the envelope so states 4 and 5 cannot appear when state 2 is a possible outcome. At larger separations, the companion inflation does not exceed the original separation ($R_\mathrm{max}<a_\mathrm{i}$), so state 2 cannot appear. Instead, we obtain systems in state 4, which may be interesting in terms of pulsar planet formation.

\section{Discussion}
\label{sec:discussion}

\subsection{How will the inflated companions be observed?}
\label{res:observation}

    So far, we have shown the state of the system after explosion by combining the results of the inflation calculation with the post-SN orbital parameters. Based on these results we will also show how such systems may be observed by considering the position on the HR diagram.

\begin{figure}
    	\begin{minipage}{1.0\hsize}
    		\centering
    		\includegraphics[width=1.0\columnwidth]{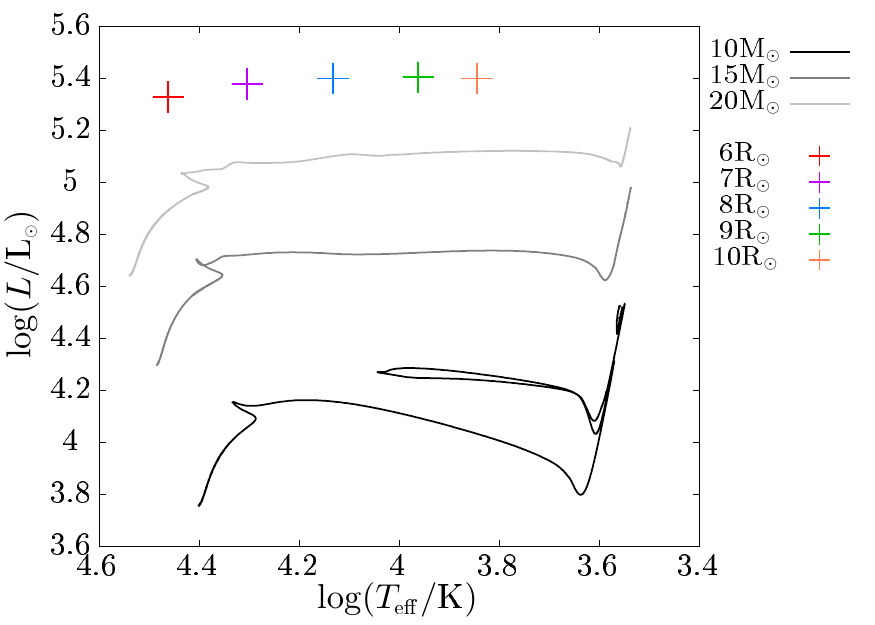}
	    	\subcaption{the envelope is not stripped}
	    	\label{HR1}
	    \end{minipage}	\\
	    \begin{minipage}{1.0\hsize}
	    	\centering
	    	\includegraphics[width=1.0\columnwidth]{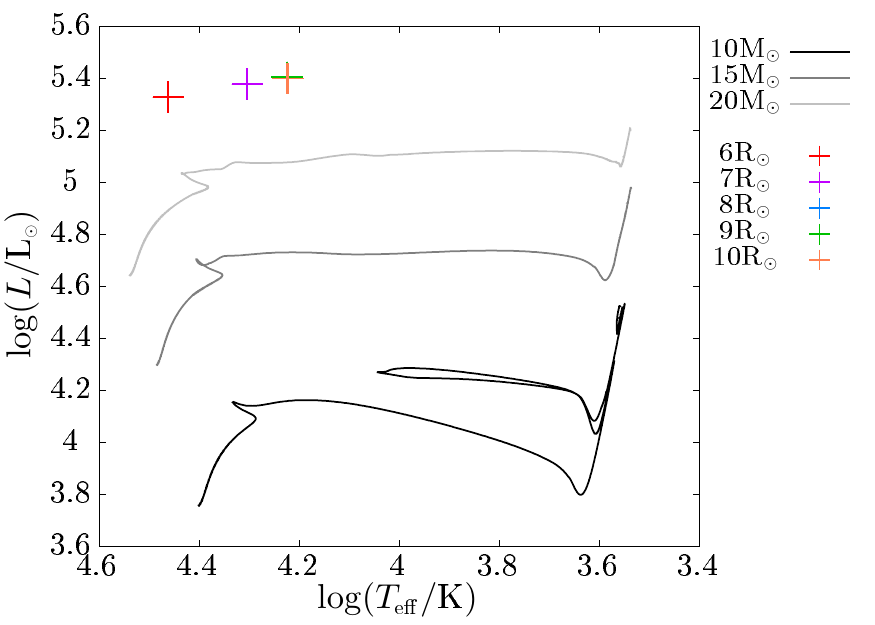}
	    	\subcaption{the envelope is stripped}
	    	\label{HR2}
	    \end{minipage}
	    \caption{HR diagram for a model $M_2=15~\msun, a_\mathrm{i}=60~\rsun, M_{\mathrm{ej}}=7.1~\msun, E_{\mathrm{expl}}=10^{51}~\mathrm{erg}$. The grey curves are single star evolution tracks for $10~\msun, 15~\msun$ and $20~\msun$ stars, and crosses represent the position of the expanded secondary due to ECI at the maximum radius. In the upper panel, the secondary envelope is not stripped by the primary, on the other hand, in the lower panel the envelope is partially stripped by the primary NS.}
    	\label{HR-diagram}
    \end{figure}

    In the upper panel of Fig.~\ref{HR-diagram}, we show the position of the secondary on the HR diagram in its maximally inflated phase. This corresponds to the end points of the tracks in Fig.~\ref{tHR} and is where the star spends most of its time before rapidly contracting back. In these figures we display models with $M_2=15~\msun, a_\mathrm{i}=60~\rsun, M_{\mathrm{ej}}=7.1~\msun, E_{\mathrm{expl}}=10^{51}~\mathrm{erg}$, with different initial secondary radii (corresponds to different ages at which the primary SN occurs). It is clear that the positions of the secondary affected by ECI are extremely far from the regular evolutionary track of a $15~\msun$ star and also even brighter than a $20~\msun$ star.
    In the lower panel, we show the same figure as Fig.~\ref{HR1}, but when the secondary expansion is suppressed by the presence of the NS. Here we assume that the NS accretes or blows away any matter that expands beyond the periastron distance of the post-SN orbit, while keeping the same high luminosity due to the continuous supply of excess energy diffusing out from the interior. This is a somewhat conservative assumption, since there could be Roche lobe overflow and the star could be constrained by its Roche lobe size instead. However, it is also possible that the NS just plunges through the envelope without significantly disturbing it. The periastron distance depends on the direction and magnitude of the kick. Here we consider the smallest possible periastron distance for a given kick velocity $V_\mathrm{kick}=100$~km~s$^{-1}$. Compared to Fig.~\ref{HR1} the larger initial radius models ($R_2=$8--10$~\rsun$) are prohibited to expand beyond the periastron distance, therefore the temperature and luminosity of these models become indistinguishable.

    We now focus on the observability of such inflated companions with various telescopes. Assuming the inflated stars emit as a black body, we estimate how far these systems could be observed with the Hubble Space Telescope (HST) and the James Webb Space Telescope (JWST). We consider three filters, \verb|HST/WFC3_UVIS1.F300X| for UV, \verb|JWST/NIRCam.F070W| for visible and \verb|JWST/NIRCam.F115W| for near-IR.

\begin{figure}
    	\begin{minipage}{1.0\hsize}
		\centering
		\includegraphics[width=1.0\columnwidth]{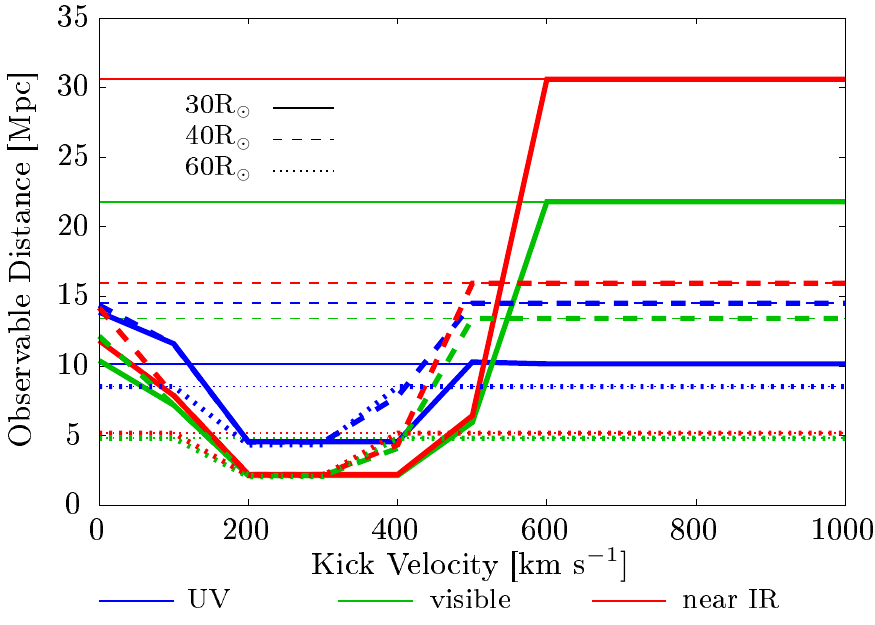}
	    \end{minipage}
	    \caption{The observable distance for the model $M_2=5~\msun, R_2=2.7~\rsun, M_{\mathrm{ej}}=5~\msun, E_{\mathrm{expl}}=10^{51}\ \mathrm{erg}$. The blue, green and red curves denote the wavelength used to observe, UV, visible and near IR respectively. Different line types show results for different orbital separations $a_\mathrm{i}=30,40,60~\rsun$. Thick curves assume that the inflation can be suppressed by the primary, while the thin curves assume no suppression.}
	    \label{observable_dis}
    \end{figure}

    We show the observable distance of such inflated companions with each wavelength in Fig.~\ref{observable_dis}. The observable distance is defined as the maximum distance where the apparent magnitude of the star is lower than a limiting magnitude, which we set to $m_\mathrm{lim}=25$~mag. The thin lines show the observable distance when there is no suppression of expansion. The thick curves are the observable distances for stars with inflation suppressed at the periastron distance of the orbit. Because the periastron distance depends on the kick direction, we show the results for the most severely suppressed cases. The true observable distance can be anywhere between the thin and thick curves depending on the kick direction. How we calculate the distances are explained in Appendix \ref{app:distance}. This model is for $M_2=5~\msun, R_2=2.7~\rsun, M_{\mathrm{ej}}=5~\msun, E_{\mathrm{expl}}=10^{51}\mathrm{erg}$. The line colours and types indicate the different observation wavelengths and initial separations. When the orbital separation is sufficiently close, we may be able to see inflated $5~\msun$ companions up to $\lesssim30$~Mpc in the IR, or $\lesssim20$~Mpc in the visible band. This will obviously be much higher for higher mass companions because the maximum luminosity is proportional to mass. Basically, any suppression of expansion will lower the detectable distance within the observational wavelengths we show here, except for some regions of the UV where suppression leads to brightening. 
It should also be noted that for the wider separation models, the inflated companion may not be observed even if it is located at sufficiently close distances ($\lesssim$5~Mpc). This is because the weaker energy models have shorter inflated timescales and therefore may shrink back while the SN is still bright.

%

\subsection{Detection probability}
\label{dis:BPS}

    Whether we can observe the inflated companions to \acp{CCSN} depend on both the maximum luminosity during the inflated phase, which is determined solely by the secondary mass, and the distance to the system. If we consider expansion suppression by the primary, it will also depend on the orbital separation. The combination of secondary masses and orbital separation is strongly dependent on the binary evolution leading to the SN. We perform rapid binary population synthesis calculations in order to obtain the distribution of secondary masses and orbital separations and estimate the detection probability of inflated secondaries.

    We use the widely used \textsc{BSE} code \citep{Hurley2002}, to evolve 1,400,000 binaries based on simplified binary evolution prescriptions. Two sets of models are computed with solar metallicity $Z=0.0134$ and sub-solar metallicity $Z=0.00134$. The initial primary masses are drawn from the Salpeter initial mass function \citep{Salpeter1955}, and the secondary mass is drawn from a flat distribution for the mass ratio $q=M_2/M_1$. The initial orbital separation is assumed to follow a log-flat distribution \citep{Abt1983}, and the initial eccentricities are set to zero for simplicity.
    The initial primary mass range is chosen to be $7<M_1/\msun<300$, and the initial mass ratio range is $0.08/M_1<q<1$ (this corresponds to an initial secondary mass range of $0.08<M_2/\msun<M_1$). The range of the initial separation is $a_\mathrm{min}/\rsun<a/\rsun<10^6$, where $a_\mathrm{min}$ is the separation at which the primary fills its Roche lobe. 
    All models are evolved up to the point of the first SN in the system. $\sim$400,000 of the binaries end up merging before exploding, so only $\sim$1,000,000 binaries experience the first SN with a companion.

    The binary physics adopted in our population synthesis calculations are as follows: When a star fills its Roche lobe, the stellar material is transferred to its companion. The mass transfer stability is determined by comparing     $\zeta_\mathrm{L}=\mathrm{d}\log R_\mathrm{L,don}/\mathrm{d}\log M_\mathrm{don}$ with $\zeta_\mathrm{ad}=(\mathrm{d}\log R_\mathrm{don}/\mathrm{d}\log M_\mathrm{don})_\mathrm{ad}$, where $M_\mathrm{don}$, $R_\mathrm{don}$ and $R_\mathrm{L,don}$ are the mass, radius and Roche lobe radius of the donor star, respectively. $\zeta_\mathrm{ad}$ depends on the stellar type, and in this study we use the following values: $-1+2M_\mathrm{don}/3M_\mathrm{env,don}$, 2.59, 6.85, 1.95 and 5.79 for the giant star with the convective envelope \citep{Hjellming1987}, main sequence, giant star with the radiative envelope \citep{Hjellming1989}, helium main sequence and helium giant star \citep{Ivanova2002, Belczynski2008}, where $M_\mathrm{env,don}$ is the envelope mass of the donor giant star. If $\zeta_\mathrm{L}>\zeta_\mathrm{ad}$, the mass transfer is dynamically unstable, and the binary system enters a common-envelope phase. We use the $\alpha\lambda$ formalism for common-envelope evolution \citep{Webbink1984}, and set the common-envelope efficiency parameter as $\alpha_\mathrm{CE}=1$. Prescriptions in \citet{Xu2010, Xu2010b} are used for the envelope binding energy parameter $\lambda_\mathrm{CE}$. We assume that the binary system coalesces after the common-envelope phase if the radius of one of the stars exceeds its Roche lobe radius \citep{Hurley2002}. If $\zeta_\mathrm{L}<\zeta_\mathrm{ad}$, the donor mass is stably transferred to its companion. We calculate the mass transfer rate using the Eq. (58) in \citet{Hurley2002}. In this study, we assume that the mass transfer is non-conservative and only half of the transferred mass can be accreted onto the companion. The non-accreted mass is assumed to be isotropically re-emitted, ejected with the specific angular momentum of the accretor. If an SN explosion occurs, we calculate the remnant mass from the carbon-oxygen core mass, adopting the `rapid' SN prescription in \citet{Fryer2012}.

    Because ECI will only be important in close binaries, we focus on the hydrogen-deficient SNe. In population synthesis codes, it is not possible to distinguish between type Ib and IIb SNe due to the way mass transfer is treated. We simply define the models with no hydrogen envelope as stripped-envelope SNe. This leaves us with $\sim$150,000 (180,000 for the sub-solar metallicity model) binaries out of the initial 1,400,000. Moreover the number of the post-common-envelope systems is $\sim$33,000 (67,000 for sub-solar metallicity) in binaries that experience stripped-envelope SNe. Most of them leave close binaries that are relevant for ECI, so as a result 77.9\% (68.8\% for sub-solar metallicity) of the close ($a_i<60~\rsun$) binaries including the stripped-envelope SN progenitors are the post-common-envelope binaries.

    For each binary in our population synthesis models, we use the secondary mass, radius and orbital separation to calculate how much the star will inflate, based on our ECI simulations\footnote{We ignore the effect of metallicity on post-ECI inflation. Instead we apply our solar metallicity ECI results even to the sub-solar metallicity population synthesis models.}. We also use the CO core mass to estimate the kick velocity (see Appendix~\ref{app:rate}). By assuming the kick is randomly directed isotropically, we calculate the probabilities of each state which the system can end up in. We then add up all the probabilities to obtain the rate of occurrence of each state over the whole stripped-envelope SN population.

    The results are summarised in Table~\ref{BPS_table2}. The most frequent case is state 5 (two stars are bound, but nothing occurs) and the probability is about 97~\%. The separation is large in these models and therefore there is little expansion due to ECI. 
    The second most frequent case is state 6 (binary disruption) with 1-2\%.
    The third most frequent case is state 2 (primary NS engulfed in the secondary envelope) with $\sim$0.8\%. State 2 and 3 is probably the most interesting among other states. There could be accretion onto the NS that cause brief X-ray transients, and/or it could suppress the expansion of the envelope. Our rate estimate implies that possibly $\sim$0.8~\% of stripped-envelope SNe may be accompanied by brief X-ray transients if the environment is sufficiently optically thin. 
    On the other hand, state 4 seems quite rare in both metallicities. This implies that pulsar planet formation through this channel may be extremely rare.

    \begin{table*}
      \centering
      \begin{tabular}{cccccccc}     \hline
        & state 1 [\%] & state 2 [\%] & state 3 [\%] & state 4 [\%] & state 5 [\%] & state 6 [\%]  \\  \hline
        Solar     & 0.27 & 0.82 & 0.064 & 2.6$\times 10^{-4}$ & 97.2 & 1.62  \\
        sub-Solar & 0.27 & 0.75 & 0.072 & 6.7$\times 10^{-4}$ & 96.8 & 2.10  \\  \hline
      \end{tabular}
      \caption{The rate that each states appear after hydrogen deficient SNe for two metallicities ($Z=1.0~Z_{\odot}, 0.1~Z_{\odot}$). The different states are defined in section \ref{ECI:mode}.}
      \label{BPS_table2}
    \end{table*}

    We estimate the expected probability at which we will be able to detect an inflated companion per stripped-envelope SN at five different wavelengths. Again, our limiting magnitude is set to $m_\mathrm{lim}=25$~mag. We calculate the detectability for five filters, \verb|HST/WFC3_UVIS1.F300X| for the U band, \verb|HST/WFC3_UVIS1.F410M| for the B band, \verb|HST/WFC3_UVIS1.J547M| for the V band, \verb|JWST/NIRCam.F070W| for the R band and \verb|JWST/NIRCam.F115W| for the I band.
    The detailed calculation methods are described in Appendix \ref{app:rate}.

    The detection probabilities are summarized in Table \ref{BPS_table}. The inflated secondaries could be observed best in the U band and the detection probability is $\sim 1.0~\%$. This means that out of all stripped-envelope SNe in the local Universe ($\leq$20~Mpc), there is $\sim$1~\% chance of detecting an inflated companion. The detection probabilities are higher for sub-solar metallicity models.
    This is likely because the stellar winds are stronger in higher metallicity, leading to wider separations and therefore less inflation after ECI. 91.6~\% (89.5~\% for sub-solar metallicity) of the detectable systems are post-common-envelope binaries in the suppressed case.
    The detectability is $\sim$3 times higher in all bands when we lift the assumption of expansion suppression. Since we do not know whether the inflation can actually be inhibited by the presence of the NS, the true detectability can be anywhere in between.

    We also note that the detectability will likely be higher if we constrain our targets to type Ib SNe. As we have mentioned before, the different types of stripped-envelope SNe cannot be distinguished in rapid binary population synthesis codes. Type Ib SNe are more severely stripped than type IIb SNe, so the typical binary separation can be expected to be closer for type Ib progenitor systems and therefore have more inflated companions. Type Ib SNe comprise roughly half of the hydrogen-deficient SNe \citep[e.g.][]{Shibbers2017}, so the detection rates will roughly be twice higher. Also, we have only derived rates for a particular set of assumptions in the population synthesis calculations. For example if we choose a smaller value for the common-envelope efficiency parameter, the typical post-common-envelope separations will be smaller and could drastically affect our rate estimates. Conversely, we may be able to use observed rates of inflated companions to constrain the binary evolution physics. We will leave this exploration to future work. 

    \begin{table}
      \centering
      \begin{tabular}{cccccc}     \hline
        Population model& U [\%] & B [\%] & V [\%] & R [\%] & I [\%]  \\  \hline
        $1.0~Z_\odot$, suppressed     & 0.90 & 0.27 & 0.39 & 0.03 & 0.0  \\
        $1.0~Z_\odot$, no suppression & 2.73 & 2.60 & 1.13 & 0.05 & 0.0 \\
        $0.1~Z_\odot$, suppressed     & 1.08 & 0.31 & 0.44 & 0.06 & 0.0  \\
        $0.1~Z_\odot$, no suppression & 3.11 & 2.95 & 1.36 & 0.11 & 0.0 \\ \hline
      \end{tabular}
      \caption{Observable probabilities of inflated companions in each band. Results are shown for two types metallicities ($Z=1.0~Z_{\odot}, 0.1~Z_{\odot}$), and with or without expansion suppression.}
      \label{BPS_table}
    \end{table}

\subsection{Implications on binary evolution}
\label{dis:implication}

    With the relations presented in Section~\ref{sec:result}, it is now possible to derive the complete set of pre-SN binary parameters from detections of inflated companions. First, we can use Eq.~(\ref{eq:Lmax}) to calculate the secondary mass from the luminosity $L$. By manipulating Eq.~(\ref{eq:Lmax}), the secondary mass can be expressed as
\begin{equation}
 \frac{M_2}{\msun}=\left(0.02+\frac{4\pi Gc\msun}{\kappa_0L}\right)^{-1},
  \label{eq:M2fromL}
\end{equation}
where $\kappa_0=1.24~\mathrm{cm}^2~\mathrm{g}^{-1}$. Note that this is only valid if the observed source is in the inflated state. Second, we can use Eq.~(\ref{eq:t_inf_fit}), to compute the injected energy using the inflated timescale
\begin{equation}
 E_\mathrm{heat}=\frac{\tau_\mathrm{inf}L}{\alpha}.
 \label{eq:E_int_ana}
\end{equation}
Then from $E_\mathrm{heat}=p E_\mathrm{expl}\tilde{\Omega}\sim p E_\mathrm{expl}(R_2/a_\mathrm{i})^2/4$, we obtain
\begin{equation}
 \frac{R_2}{a_\mathrm{i}}\sim\sqrt{\frac{4\tau_\mathrm{inf}L}{\alpha p E_\mathrm{expl}}},
\end{equation}
where $p\sim0.08-0.10$ and $\alpha\sim0.18$. Assuming that the explosion energy is constrained from other diagnostics such as the SN light curve, these relations allow us to explicitly compute the pre-SN binary parameters. Obtaining the required information such as the companion luminosity $L$ and the inflated timescale $\tau_\mathrm{inf}$ requires multi-band photometry over multiple epochs. Single band photometry will only allow us to constrain the luminosity very loosely. We can also only place lower limits on the duration of expansion unless the contraction back from the inflated state is observed.

There is also the possibility that the surface of the companion is polluted with heavy elements captured from the slower parts of the SN ejecta. The amount of pollution is usually very small \citep[$\lesssim10^{-3}~\msun$;][]{Liu2015, Hirai2018}, and it is also strongly dependent on the orbital separation. The accreted elements will be quickly mixed through the surface convective layers of the inflated envelope. It is not clear how observable these chemical features would be, but it may provide an additional constraint on the orbital separation.


    Extracting the full set of pre-SN binary parameters for an observed system may have extremely important implications for binary evolution theory. The systems that have sufficiently inflated companions are the systems with close orbital separations, which mostly have experienced common-envelope evolution. Despite its great importance in stellar astrophysics, common-envelope phases are one of the least understood phases in binary evolution \citep[see][for a review on common-envelope evolution]{Ivanova2013}. Of particular importance is the orbital separation after common-envelope phases, which determines how the binary evolves thereafter. Because the pre-SN orbital separations are closely related to the post-common-envelope separation, the detections of inflated companions may provide the best constraints on the outcome of common-envelope phases to date. This will give insight into the long sought common-envelope efficiency parameter ($\alpha_\textsc{ce}$), which has so far been difficult to calibrate from observations especially in the massive star regime. For the systems that only experienced stable mass transfer, the pre-SN orbital separation would be determined by how conservative the mass transfer is, and the mode of angular momentum loss. Observations of inflated companions can help probe these effects too.

\subsection{Application to SN~2006jc}
\label{dis:SN2006jc}

    There are still only a limited number of SNe that have detections of companion stars. It is also crucial to obtain multi-epoch observations of the companion in order to constrain the inflated timescale, which can be used to estimate the pre-SN binary parameters. One case that satisfies these requirements is SN~2006jc \citep[]{Foley2007, Pastorello2007, Maund2016, Sun2019}.

\begin{figure}
	    \begin{minipage}{1.0\hsize}
		    \centering
		    \includegraphics[width=1.0\columnwidth]{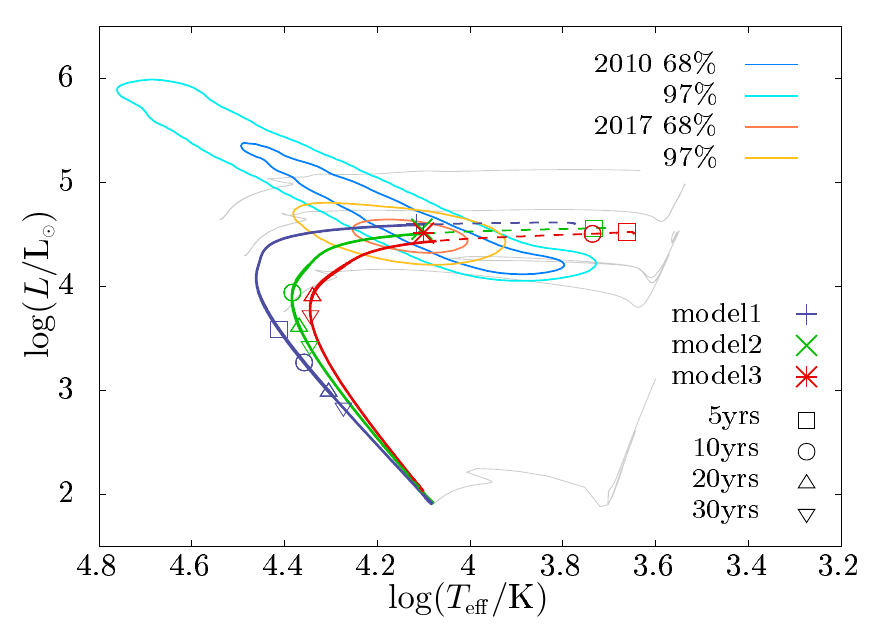}
	    \end{minipage}
	    \caption{Evolutionary tracks of post-ECI models with $M_2=3~\msun, R_2=2~\rsun, a_\mathrm{i}=40~\rsun, M_{\mathrm{ej}}=2~\msun$ on the HR diagram. The various cross symbols mark the locations if the envelope expansion is suppressed by the primary. The three models have different explosion energies, $E_{\mathrm{expl}}=1, 5, 10 \times 10^{51}\ \mathrm{erg}$ respectively. The three solid curves are the evolutionary tracks in the suppressed model, and the dashed curves are the evolutionary tracks without the suppression. The square, circle and triangles are time stamps at $5, 10, 20$ and $30$ years after SN explosion, and these are also for the non-suppressed model.
	    The contours show the probability of SN 2006jc companion's position from 2010 and 2017 observations \citep[]{Sun2019}.}
	    \label{HR_SN2006jc}
    \end{figure}

The SN site was examined in 2010 and 2017, $\sim$3 and 10 years after the explosion, both showing fluxes consistent with each other. From multi-band photometry of the post-SN source, the temperature and luminosity of the companion to SN~2006jc were constrained and is consistent with being a Hertzsprung gap star of mass $M_2\sim$10--15~$\msun$ \citep[Fig.~\ref{HR_SN2006jc},][]{Maund2016, Sun2019}. Although it is feasible, it is also a rare case because the Hertzsprung gap phases of massive stars are very short-lived. \citet{Sun2019} investigated whether this could instead be a low-mass star inflated due to ECI, and estimated the binary parameters based on this scenario. However, there was one issue with the scenario. For models that can inflate for more than 10 years, the maximum radius is also quite large and tend to lie closer to the red supergiant regime, while the observed source was bluer.

Here we will bring in the additional assumption that the stellar expansion could have been inhibited by the presence of the NS. This would allow the inflated star to appear hotter, while having the high luminosity and long inflated timescales. Based on this scenario, we can already constrain the pre-SN binary parameters using the analytical formulae we derived above. The luminosity of the source is $\log(L/\lsun)\sim4.5$. From Eq.~\ref{eq:M2fromL}, the secondary mass should be $\sim3~\msun$. Because the companion is inflated for $\gtrsim$10~yr, Eq.~(\ref{eq:E_int_ana}) gives $E_\mathrm{int}\gtrsim3\times10^{48}$~erg. The estimated temperature of the companion corresponds to a star with a radius $\sim40~\rsun$. Assuming that the radius is constrained by the periastron distance of the post-SN orbit, the pre-SN separation should be $a_\mathrm{i}\gtrsim40~\rsun$. This leaves us with only the explosion energy as a free parameter.


    The contours in Fig.~\ref{HR_SN2006jc} show the HR diagram constraints of the SN~2006jc companion at two epochs. We overplot three post-ECI models run with parameters summarized in Table~\ref{SNmodel}. By assuming a suppressed expansion due to the presence of the primary NS, all models lie within the observational constraints. The expanded timescale is from a few to tens of years depending on the explosion energy. From the follow-up observations, SN 2006jc has a flat light curve for at least ten years. Therefore if the companion of SN 2006jc is indeed a post-ECI star where its expansion is suppressed by the primary NS, the model with relatively high explosion energy (model 3, $E_{\mathrm{expl}}\gtrsim 10^{52}\mathrm{erg}$) is compatible with the observations. This is consistent with some theoretical estimates from the SN light curve \citep[]{Tominaga2008}.

    \begin{table*}
      \centering
      \begin{tabular}{ccccccc}     \hline
        & $M_2 [\msun]$ & $R_2 [\rsun]$ & $a_\mathrm{i} [\rsun]$ & $M_{\mathrm{ej}} [\msun]$ & $E_{\mathrm{expl}} [\mathrm{erg}]$ & timescale [yr]  \\  \hline
        model 1    & 3 & 2.0 & 40 & 2 & $10^{51}$        & 1.5 \\
        model 2    & 3 & 2.0 & 40 & 2 & $5\times10^{51}$ & 7.0 \\
        model 3    & 3 & 2.0 & 40 & 2 & $10^{52}$        & 13.0 \\
        Sun et al. & 4 & 2.5 & 57 & 2 & $10^{52}$        & - \\  \hline
      \end{tabular}
      \caption{Simulation parameters for the models plotted in Fig.~\ref{HR_SN2006jc}.}
      \label{SNmodel}
    \end{table*}

Within this scenario, we expect the companion to contract back to its original radius within a few years. For example in model 3, the temperature and luminosity drop out of the contours $\sim$15 years after the SN (red triangle). Therefore, it will be vital to carry out an additional observation of the SN site some time soon to further constrain the model.

    In \citet{Sun2019}, the assumed binary parameters were $a_\mathrm{i}=57~\rsun, M_2=4~\msun, R_2=2.5~\rsun, M_{\mathrm{ej}}=5~\msun, E_{\mathrm{expl}}=10^{52}\mathrm{erg}$. Compared with their results, the system of SN 2006jc could be a closer binary, if the stellar expansion can indeed be inhibited by the primary.

\section{Summary}
\label{sec:summary}

	We systematically study the outcome of SNe in binary systems whose companions are MS stars. We carry out 720 1D stellar evolution calculations of companion stars for 10,000 years after being impacted by \ac{SN} ejecta. This is done using the 1D stellar evlution code MESA by injecting energy into the star based on the formulation of \citetalias{Hirai2018}. We explore a wide parameter space, varying the secondary mass, secondary radius (or age), orbital separation, ejecta mass and explosion energy. 

    The companion star is driven out of thermal equilibrium and immediately expands its envelope for several years before rapidly contracting back to its original size. Most of the inflated companions exist in the Hertzsprung gap of the HR diagram, and are much more luminous than its regular luminosity. The maximum luminosities are determined by the Eddington luminosity computed with Fe opacity. We derive an empirical formula for the relevant opacity as a function of the secondary mass, which can be used to relate the post-SN luminosity and mass of the companion. The duration of expansion is tightly correlated with the injected energy. If we can observe inflated companions in the future, the luminosity and inflated timescales can be used to determine the full set of pre-SN binary parameters. Such detections may have extremely important implications for understanding binary evolution physics, especially on common-envelope evolution.

    We discuss the various different states the binary can take after the SN. We identify 6 different states (written in section \ref{ECI:mode}) which are determined by the kick velocity and direction. For example, part of the inflated envelope may be transferred to the new-born NS, possibly leading to the formation of pulsar planets. As another possible outcome, the NS may directly plunge into the inflated part of the envelope, which could lead to short X-ray transients and/or the accretion feedback may blow away the outer parts of the envelope. This may be particularly interesting as it will affect the appearance of the secondary star, making it smaller in size and therefore hotter in temperature.


    We also performed simple binary population synthesis calculations to estimate the observability of companions. We estimate that 1--3~\% of hydrogen deficient SNe within <20~Mpc may have observable inflated companions which will be most detectable in the U band.

    Finally, we apply our model to the observed companion of SN~2006jc which is a type Ibn SN. Using our analytic formulae, we estimate the binary parameters to be $M_2=3~\msun, R_2=2~\rsun, a_\mathrm{i}=40~\rsun$, and the explosion energy should be relatively large ($E_\mathrm{expl}\sim10^{52}$~erg). We apply a new assumption that the inflation could be suppressed due to the presence of the NS. Compared with \cite{Sun2019}, the progenitor of SN 2006jc could be a closer binary. The inflated timescale can strongly vary depending on the explosion energy. Therefore, if future observations reveal that the companion has faded, it will provide additional strong constraints on the explosion energy.

\section*{Acknowledgements}
The authors thank the anonymous referee for constructive comments that improved the manuscript. The authors thank Ning-Chen Sun for providing data of the HR diagram constraints for SN~2006jc. MO thanks Keiichi Maeda and the COMPAS team for fruitful discussions during the development of the project.

\section*{Data Availability}
    The data underlying this article will be shared on reasonable request to the corresponding author.



\bibliographystyle{mnras}
\bibliography{example} 




\appendix

\section{Post-SN orbital variables}
\label{app:orbit}

    The orbital variables after \ac{SN} are computed based on an approximation that the explosion is impulsive.
    First, we introduce velocities related to orbital motion. As we assume an initially circular orbit, the orbital velocity just before the explosion $V_{\mathrm{i}}$ is
    \begin{equation}
    	V_{\mathrm{i}} = \left( \frac{G(M_1+M_2)}{a_\mathrm{i}} \right)^{1/2},
    \end{equation}
    where $G$ is the gravitational constant and $M_1$ the primary mass, $a_\mathrm{i}$ is the initial orbital separation. We set $M_1=M_{\mathrm{ej}}+m_c$, where $m_c$ is the primary core mass and equal to the remnant NS mass, and we use $m_c=1.4~\msun$ for all of our models. We also consider the natal kick imparted to the remnant of the exploding star which is expressed as
    \begin{align}
    	V_{\mathrm{kick,x}} &= V_{\mathrm{kick}} \sin \theta \cos \phi,	\\
        V_{\mathrm{kick,y}} &= V_{\mathrm{kick}} \cos \theta,	\\
        V_{\mathrm{kick,z}} &= V_{\mathrm{kick}} \sin \theta \sin \phi,
    \end{align}
    where $V_\mathrm{kick}$ is the magnitude of the kick.
    The coordinate angles are defined in Fig.~\ref{setup}.
    The impact velocity of the secondary is also taken into account and estimated analytically based on the formula \citepalias{Hirai2018}
    \begin{equation}
    	V_{\mathrm{im}} = \eta\left( 2E_{\mathrm{expl}}M_{\mathrm{ej}} \right)^{1/2}\frac{\tilde{\Omega}}{M_2},
    \end{equation}
    where $\eta$ and $\tilde{\Omega}$ are the efficiency of momentum transfer ($\sim1/3$) and the fractional solid angle to the companion ($=\Omega/4\pi$), and the expression in the brackets is a rough estimate of the total momentum of the ejecta.

    The separation and eccentricity of the orbit after SN explosion $a_\mathrm{f}, e$ is computed with the following formula \citep{Konstantin2014}
    \begin{align}
     a_\mathrm{f} &= a_\mathrm{i} \left[ 2- X \frac{(V_{\mathrm{kick,x}}+V_{\mathrm{i}})^2+(V_{\mathrm{kick,y}}+V_{\mathrm{im}})^2+V_{\mathrm{kick,z}}^2}{V_{\mathrm{i}}^2} \right]^{-1},\\
     e &= \left( 1-X\frac{a_\mathrm{i}}{a_\mathrm{f}}\frac{(V_{\mathrm{kick,y}}+V_{\mathrm{im}})^2+V_{\mathrm{kick,z}}^2}{V_{\mathrm{i}}^2} \right)^{1/2},
    \end{align}
    where $X=(M_1+M_2)/(m_c+M_2)$ is the mass ratio of the total masses before and after explosion.

    When $e>1$, the binary is disrupted. If the binary survives, the orbital period $P_{\mathrm{orb}}$ is
    \begin{equation}
    	P_{\mathrm{orb}} = 2\pi \left( \frac{a_\mathrm{f}^3}{G(m_c+M_2)} \right)^{1/2}.
    \end{equation}

    The Roche lobe radius \citep{Eggleton1983} of the secondary is
    \begin{equation}
    	R_{\mathrm{rl}} = \frac{0.49q^{2/3}}{0.6q^{2/3}+\ln(1+q^{1/3})} a_\mathrm{i},
    \end{equation}
    where $q=M_2/M_1$ is the mass ratio. We use $R_{\mathrm{rl}}$ to evaluate whether Roche-lobe overflow commences or not.

\section{Observable distance}
\label{app:distance}

    To calculate the observable distance of inflated secondary, we approximate the emission from the secondary as a black-body. The radiation intensity $B(T,\lambda)$ is
    \begin{equation}
    	B(T,\lambda) = \frac{2hc^2}{\lambda^5} \left[ \exp\left(\cfrac{hc}{kT\lambda}\right)-1 \right]^{-1},
    \end{equation}
    where $h,c,k,\lambda$ and $T$ are the Planck constant, the speed of light, the Boltzmann constant, emission wavelength and temperature respectively. Using the radiation intensity we obtain the specific flux $F_{\lambda}(10\mathrm{pc})$ for wavelength $\lambda$ at $10\mathrm{pc}$ away in order to seek the absolute magnitude
    \begin{equation}
    	F_{\lambda}(10\mathrm{pc}) = \pi B(T,\lambda) \left( \frac{R}{10\mathrm{pc}} \right)^2,
	 \label{eq:abs_flux}
    \end{equation}
    where $R$ is the stellar radius. Then the absolute magnitude at a given wavelength $\lambda$ follows from
    \begin{equation}
    	M_{\mathrm{abs}}=2.5 \log_{10}\left[ \frac{F_{\mathrm{zero\ point}}}{F_{\lambda}(10\mathrm{pc})} \right],
	 \label{eq:mabs}
    \end{equation}
    where $F_{\mathrm{zero\ point}}$ is the zero point flux for each filter of the telescopes. We then compute the maximum observable distance for a given limiting magnitude from
    \begin{equation}
    	d=10^{\left(m_{\mathrm{lim}}-M_{\mathrm{abs}}+5-A(\lambda)\right)/5},
	 \label{eq:max_dis}
    \end{equation}
    where $m_{\mathrm{lim}}, A(\lambda)$ are limiting magnitude and interstellar extinction. We set the limiting magnitude to 25 mag for all wavelengths. We use laws in \citet{Cardelli1989} for the interstellar extinction that depends on the wavelength. As typical extinction values for SN Ibc, we use $E(B-V)=0.4$ and $R_{\mathrm{V}}=3.1$ according to \citet{Drout2011}.

\section{Rate calculation}
\label{app:rate}

    To calculate the observable rate, we make full use of our binary population synthesis data and our post-ECI inflation simulation results.
    First, we assign a maximum expansion radius ($R_\mathrm{2,max}$) and maximum luminosity ($L_\mathrm{max}$) to each binary in the population synthesis data set by interpolating between our MESA inflated star models with the closest binary parameters $M_2, R_2, a$.
    We also randomly assign a distance to each binary assuming they are uniformly distributed within 20~Mpc from the observer.

    For a given binary model and observation wavelength $\lambda$, we can calculate the minimum required absolute flux $F_\lambda$(10pc) such that the apparent magnitude is brighter than the limiting magnitude $m_\mathrm{lim}=25$~mag, using Eqs.~(\ref{eq:mabs})--(\ref{eq:max_dis}). Then we plug this into Eq.~(\ref{eq:abs_flux}) and solve for $T$, which gives the minimum temperature required for the star to have a specific flux above the detection limit ($\equiv T_\mathrm{min}$). Finally, by assuming the star is emitting as a black body, the minimum observable radius of the star is given by
    \begin{equation}
	    R_{\mathrm{min,obs}}=\left( \frac{L_{\mathrm{max}}}{4\pi\sigma T_{\mathrm{min}}^4} \right)^{1/2}.
    \end{equation}
    Whenever the maximum expansion radius of the star exceeds the minimum observable radius ($R_\mathrm{2,max}\geq R_\mathrm{min,obs}$), the star is ``detectable''. For the case where we do not take into account any expansion suppression, we simply count the number of detectable systems and divide by the total number of stripped-envelope SNe to obtain the detection rate.

\begin{figure}
	    \begin{minipage}{1.0\hsize}
		    \centering
		    \includegraphics[width=1.0\columnwidth]{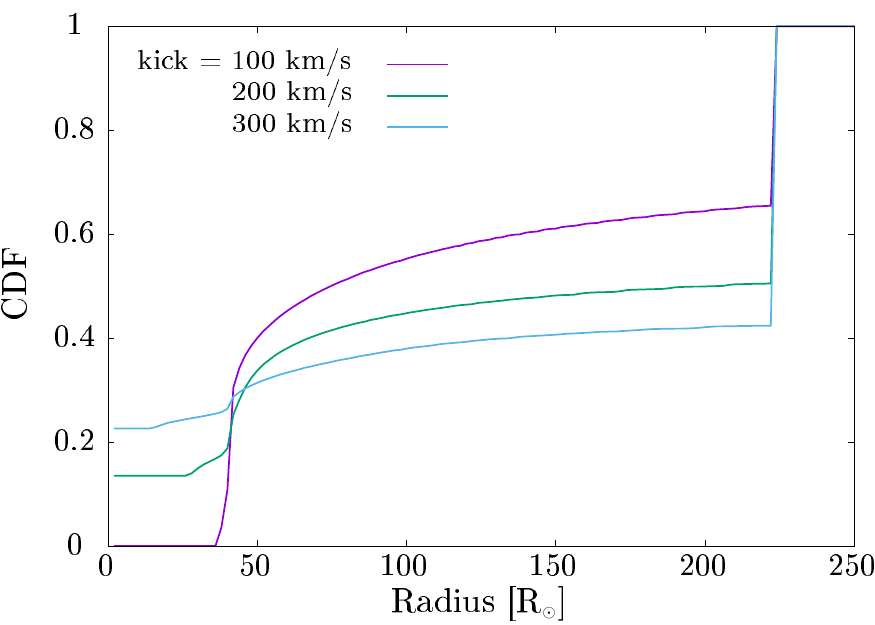}
	    \end{minipage}
	    \caption{Cumulative probability distributions of the post-ECI radius for the model $M_2=15~\msun, R_2=8~\rsun, a_\mathrm{i}=40~\rsun, M_{\mathrm{ej}}=7.1~\msun, E_{\mathrm{expl}}=10^{51}\ \mathrm{erg}$, assuming that the radius expansion is limited to the periastron distance. Different colours denote different kick velocities $V_{\mathrm{kick}}=100, 200$ and $300~\mathrm{km~s}^{-1}$.}
	    \label{prob_rad}
    \end{figure}

    For the case where we consider expansion suppression by the primary, we also take into account the effect of the NS kick on the orbit.
    For the kick magnitudes, we use the fitting formula presented in \citet{Alejandro2018}, that relates the velocity to the CO core mass $m_\mathrm{CO}$
    \begin{equation}
      V_{\mathrm{kick}} = \left\{
      \begin{array}{cc}
        35+1000(m_{\mathrm{CO}}-1.372), & 1.372\leq m_{\mathrm{CO}} < 1.49  \\
        90+650(m_{\mathrm{CO}}-1.49),   & 1.49 \leq m_{\mathrm{CO}} < 1.65  \\
        100+175(m_{\mathrm{CO}}-1.65),  & 1.65 \leq m_{\mathrm{CO}} < 2.4   \\
        200+550(m_{\mathrm{CO}}-2.4),   & 2.4  \leq m_{\mathrm{CO}} < 3.2   \\
        80+120(m_{\mathrm{CO}}-3.2),    & 3.2  \leq m_{\mathrm{CO}} < 3.6   \\
        350+50(m_{\mathrm{CO}}-4.05),   & 4.05 \leq m_{\mathrm{CO}} < 4.6   \\
        275-300(m_{\mathrm{CO}}-5.7),   & 5.7  \leq m_{\mathrm{CO}} < 6.0
      \end{array}
      \right.
      \label{Vkick}
    \end{equation}
    This gives velocities ranging from $0$ to $700~\mathrm{km~s^{-1}}$. Whenever $m_{\mathrm{CO}}$ falls outside the mass range, we assume a BH is formed and therefore there was no strong SN.
    With the given kick velocity, we calculate a cumulative probability distribution of the periastron distance assuming the kick direction is uniformly distributed over all solid angles. This corresponds to the probability of the post-SN radius of the star within our scenario. In Fig.~\ref{prob_rad} we show some sample cumulative probability distributions of the radius for a model $M_2=15~\msun, R_2=8~\rsun, a_\mathrm{i}=40~\rsun, M_{\mathrm{ej}}=7.1~\msun, E_{\mathrm{expl}}=10^{51}\ \mathrm{erg}$. Kick velocities are $V_{\mathrm{kick}}=100, 200$ and $300~\mathrm{km~s}^{-1}$. The maximum radius is about $R_\mathrm{2,max}\sim220~\rsun$ in this model. The distributions of 200 and 300 $\mathrm{km~s}^{-1}$ do not start from 0 because we set the radius to zero whenever the system ends in state 1 (stellar collision).

    From the cumulative distribution function of radius, we can compute the probability of the star to have a radius exceeding $R_\mathrm{min,obs}$, which we denote as $P_{\mathrm{obs},i}$. This means the system $i$ has a probability $P_{\mathrm{obs},i}$ of being observable. Then the detection probability for the whole population is computed from
    \begin{equation}
	    P_{\mathrm{obs}}=\frac{1}{N}\sum_i P_{\mathrm{obs,i}},
    \end{equation}
    where $N$ is the total number of stripped-envelope SNe.




\bsp	
\label{lastpage}
\end{document}